\documentclass[fleqn,usenatbib]{mnras}

\usepackage{newtxtext,newtxmath}

\usepackage[T1]{fontenc}

\DeclareRobustCommand{\VAN}[3]{#2}
\let\VANthebibliography\thebibliography
\def\thebibliography{\DeclareRobustCommand{\VAN}[3]{##3}\VANthebibliography}

\usepackage{graphicx,amsmath}

\newcommand{\st}{{\it St}}

\newcommand{\abin}{a_{\rm bin}}
\newcommand{\Tbin}{T_{\rm bin}}

\newcommand{\racc}{r_{\rm acc}}

\newcommand{\cs}{c_{\rm s}}

\newcommand{\alphagz}{\alpha_{g,z}}
\newcommand{\deltavgz}{\delta v_{g,z}^2}
\newcommand{\mdeltavgz}{\sqrt{\overline{\deltavgz}}}
\newcommand{\ddz}{D_{d,z}}
\newcommand{\dgz}{D_{g,z}}

\title[ Pebble settling in turbulent circumbinary discs]{Vertical settling of pebbles in turbulent circumbinary discs and the in situ formation of circumbinary planets}
\author[A. Pierens,  R.~P. Nelson, C.~P. McNally,]{Arnaud Pierens $^{1}$, Richard P. Nelson$^{2}$, Colin P. McNally $^{2}$ \\
$^1$ Universit\'e de Bordeaux, Observatoire Aquitain des Sciences de l'Univers,
    All\'ee Geoffroy St. Hilaire, 33165 Pessac, France\\
 $^2$  Astronomy Unit, Queen Mary University of London, Mile End Road, London, E1 4NS, UK}

\pagerange{\pageref{firstpage}--\pageref{lastpage}} 

\def\LaTeX{L\kern-.36em\raise.3ex\hbox{a}\kern-.15em
    T\kern-.1667em\lower.7ex\hbox{E}\kern-.125emX}
\begin{document}
\label{firstpage}
\maketitle
\begin{abstract}
The inner-most regions of circumbinary discs are unstable to a parametric instability whose non-linear evolution is hydrodynamical turbulence. This results in significant particle stirring, impacting on planetary growth processes such as the streaming instability or pebble accretion. In this paper, we present the results of three-dimensional, inviscid global hydrodynamical simulations of circumbinary discs with embedded particles of 1 cm size. Hydrodynamical turbulence develops in the disc, and we examine the effect of the particle back-reaction on vertical dust. We find  that higher solid-to-gas ratios lead to smaller gas vertical velocity fluctuations, and therefore to smaller dust scale heights. For a metallicity $Z=0.1$, the dust scale height near the edge of the tidally-truncated cavity is $\sim 80\%$ of the gas scale height, such that growing a Ceres-mass object to a 10 $M_\oplus$ core via pebble accretion would take longer than the disc lifetime.  Collision velocities for small particles are also higher than the critical velocity for fragmentation, which precludes grain growth and the possibility of forming a massive planetesimal seed for pebble accretion. At larger distances from the binary, turbulence is weak enough to enable not only efficient pebble accretion but also grain growth to sizes required to trigger the streaming instability. In these regions, an in-situ formation scenario of circumbinary planets involving the streaming instability to form a massive planetesimal followed by pebble accretion onto this core is viable. In that case, planetary migration has to be invoked to explain the presence of circumbinary planets at their observed locations.
\end{abstract}

\section{Introduction}

Among the 11 circumbinary planets discovered by the Kepler mission,  all but one orbit close to the dynamical instability limit, in a region where the structure of the circumbinary disc was expected to be highly perturbed during the planet formation stage. Circumbinary discs are therefore challenging environments for forming planets, such that they can provide a testbed that can constrain competing theories of planet formation.   By extension, assuming that planets form similarly around binaries and single stars,  testing models of planet formation around binaries can also provide hints about planet formation processes around single stars. 

Two competing ideas have been proposed for forming circumbinary planets. The migration hypothesis assumes that circumbinary planets formed in the outer regions of the circumbinary disc, in a more accretion friendly environment, and then migrated inwards to the location where they are currently observed. Results of hydrodynamical simulations of the migration of planets in circumbinary discs (Nelson 2003; Pierens \& Nelson 2007, 2008a,b, 2013; Kley \& Haghighipour 2014, 2015; Penzlin et al 2021), which suggest migration stalling at the edge of the central cavity, are consistent with the observed locations of circumbinary planets. Moreover, the fact that mainly sub-Jovian circumbinary planets have been detected so far, and the only Jovian mass circumbinary planet that has been detected, Kepler-1647b (Kostov et al. 2016), has a long period ($\sim 1100$ days), is also in agreement with the expectations of hydrodynamical simulations (Pierens \& Nelson 2008), and hence with the migration scenario.

 Although the migration hypothesis is an appealing scenario, it can not be excluded that circumbinary planets may be formed in-situ,  either through planetesimal or pebble accretion. Regarding planetesimal accretion, previous studies have however shown that due to the tidal influence of the binary,  planetesimals tend to acquire eccentricities large enough for the collisions to become erosive (Meschiari 2012 a,b; Paardekooper et al. 2012; Bromley \& Kenyon 2015).
 
The efficiency of pebble accretion as a means of forming circumbinary planets in situ has been examined by Pierens et al. (2020; hereafter Paper I).  They showed  that in circumbinary discs, the time needed for a Ceres-mass object to grow to the pebble isolation mass through pebble accretion is of the same order as the disc lifetime. Inefficient pebble accretion occurs because dust grains are puffed up and form a layer with finite thickness $H_d\gtrsim 0.1H$, with $H$ being the gas scale height,  resulting in a decrease of the pebble accretion efficiency by a factor of $H_d/\racc$, where $\racc$ is the accretion radius. This is a direct consequence of the turbulence operating in the disc, resulting from the parametric instability involving the resonant interaction between inertial-gravity waves and an eccentric mode in the disc (Papaloizou 2005; Barker \& Ogilvie 2014). Since circumbinary discs become eccentric through interaction with the central binary, they are good candidates for generating turbulence through the eccentric parametric instability. In paper I, we found that the resulting turbulence  transports angular momentum outwards with an effective viscous stress parameter $\alpha \sim 5\times 10^{-3}$, and the vertical velocity fluctuations are a few percent of the sound speed. In this initial study, we neglected the effects of the back-reaction from the grain particles onto the gas, and the primary aim of this paper to examine the effect of including the back-reaction. It is expected that turbulence will be weakened when dust feedback is included, as this decreases the sound speed of the dust-gas mixture and increases its inertia  (Lin \& Youdin 2017).  When turbulence originates from the Vertical Shear Instability (VSI: Nelson et al. 2013), Lin (2019) indeed found that the particle back-reaction favours dust settling against turbulence when the dust-to-gas ratio is larger than the nominal value of $Z=0.01$, because a vertical gradient in the dust-to-gas ratio induces a buoyancy force that  tends to stabilize the VSI (Lin \& Youdin 2017). In this work we examine how varying $Z$ influences the turbulence in the inner regions of a circumbinary disc, and what the implications are for forming the observed circumbinary planets in situ.

This paper is organised as follows. In Sect. 2, we describe the hydrodynamical model and numerical setup. In Sect. 3, we present the results of our 3D simulations of the vertical settling of pebbles in turbulent circumbinary discs, and discuss the impact on the turbulence and on planet formation when the particle back-reaction onto the gas is included. Finally, we draw our conclusions in Sect. 4.

\section{The hydrodynamic model}
\subsection{Governing equations}
\subsubsection{Gas component}
We solve the hydrodynamical equations for the conservation of mass,  momentum, and internal energy in spherical coordinates $(r,\theta,\phi)$ (radial, polar, azimuthal), with the origin of the frame located at the centre of mass of the binary. For the gas component, these equations read:
 \begin{equation}
 \frac{\partial \rho}{\partial t}+{\bf \nabla}\cdot(\rho {\bf v})=0,
 \end{equation}
 \begin{equation}
 \rho\left(\frac{\partial {\bf v}}{\partial t}+{\bf v}\cdot{\bf \nabla} {\bf v}\right)=-{\bf \nabla} P-\rho {\bf \nabla} \Phi-\rho_d{\bf f_{d}},
 \end{equation}
 \begin{equation}
  \frac{\partial e}{\partial t}+{\bf \nabla}\cdot(e {\bf v})=-(\gamma-1) e {\bf \nabla}\cdot {\bf v}+{\cal Q}_{cool} ,
  \label{eq:energy}
 \end{equation}
where $\rho$ is the density, $P$ the pressure, ${\bf v}$ the velocity,   $e$ the internal energy,   $\gamma$ the adiabatic index, which is set to 
$\gamma=1.4$.  In the previous equation,   ${\bf f_{d}}$ represents the mutual drag force between the gas and the solids, and $\Phi$  is the  gravitational potential of the binary, whose parameters match those of Kepler-16 (Doyle et al. 2011; see Table 1). We note that $\Phi$ includes the contribution from the indirect term, However, we do not include the gravitational back-reaction from the disc onto the binary, so the binary orbit remains fixed.    

\begin{figure}
\centering
\includegraphics[width=\columnwidth]{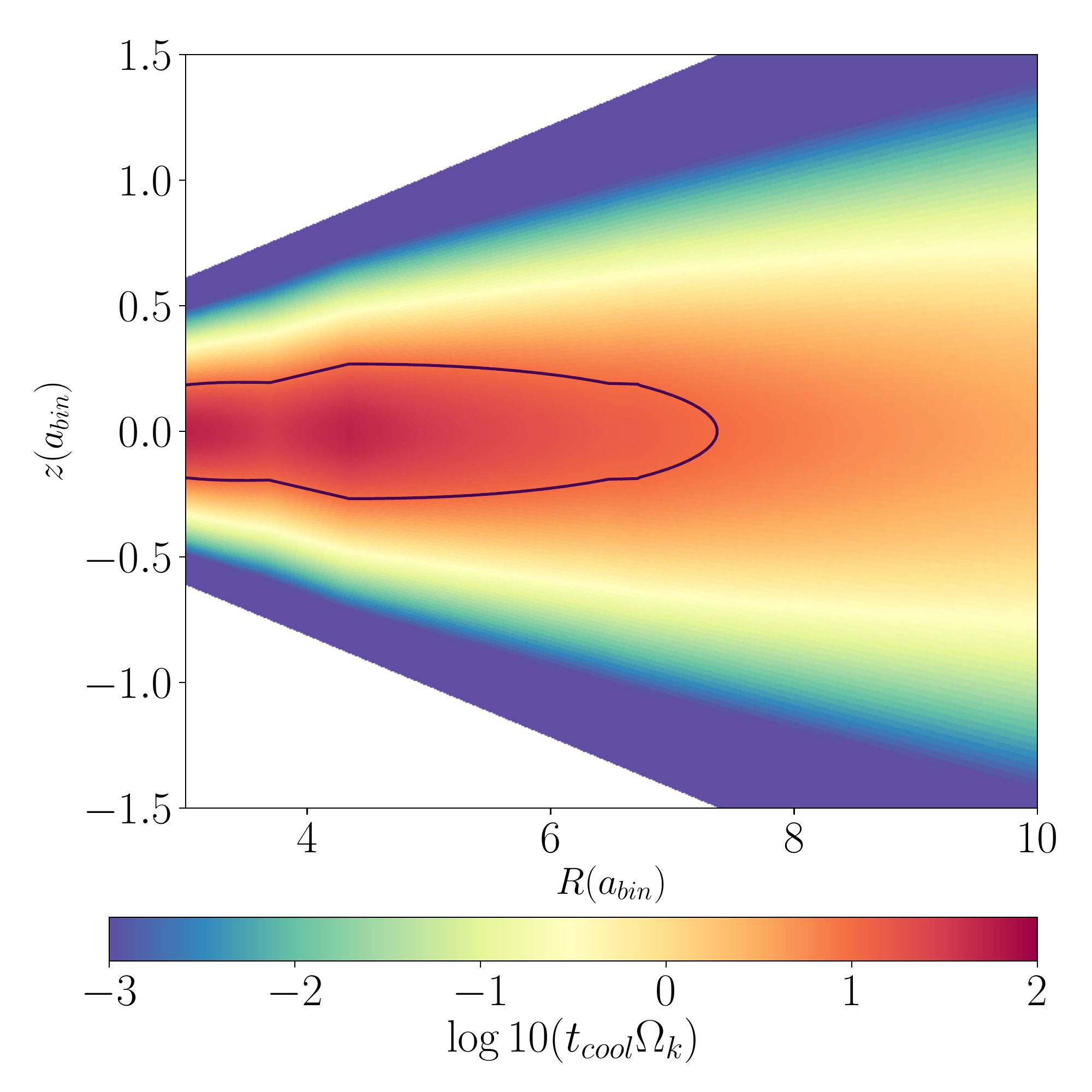}
\caption{Two-dimensional R-Z distribution of  the dimensionless cooling time for the initial disc model. The black line corresponds to the limit   where $t_{cool}\Omega=1$.}
\label{fig:disc0}
\end{figure}

Compared to our previous work, we make use of a more realistic cooling scheme that assumes that the cooling rate ${\cal Q}_{\rm cool}$ in Eq.  \ref{eq:energy} is given by:
\begin{equation}
{\cal Q}_{\rm cool}=-\rho c_v\frac{T-T_0}{t_{cool}},
\end{equation}
where $c_v$ is the specific heat capacity at constant volume,  $T_0$ the background disc temperature and $t_{cool}$ the cooling timescale. $t_{cool}$ is calculated every time step by considering 
the timescale for radiative loss of energy from a gaussian sphere of scale height corresponding to the gas scale height $H$. It is given by:
\begin{equation}
t_{cool}=\frac{\rho c_v H \tau_{eff}}{3\sigma T^3},
\end{equation}
where $\sigma$ is the Stefan-Boltzmann constant, and where $\tau_{eff}$ is the effective optical depth which is given by:
\begin{equation}
\tau_{eff}=\frac{3}{8}\tau+\frac{\sqrt 3}{4}+\frac{1}{4\tau}.
\end{equation}
Following Bae et al. (2016), the optical depth $\tau$ is calculated as $1/\tau=1/\tau_{upper}+1/\tau_{lower}$ with:
\begin{equation}
\tau_{upper}=\int_z^{z_{max}} \rho(z')\kappa(z')dz'
\end{equation}
and
\begin{equation}
\tau_{lower}=\int_{z_{in}}^z \rho(z')\kappa(z')dz',
\end{equation}
where the opacity $\kappa$ computed using the Rosseland mean opacity of Zhu et al. (2009). We show in the left panel of Fig. \ref{fig:disc0} the cooling time for our initial 
disc model which has mass equivalent to the MMSN (see Sect. \ref{sec:init}). We see that in the disc midplane $t_{cool}\Omega \in [1, 100]$ such that we expect from the results of Paper I that this region is prone to the eccentricity-induced parametric instability.

\begin{table}
\caption{Binary parameters for the Kepler-16 system (from Doyle et al. 2011).}              
\label{table1}      
\centering                                      
\begin{tabular}{c c}          
\hline\hline                        
Parameter label  & Kepler-16    \\ 
\hline 
$M_1(M_\odot)$ & $0.69$\\
$M_2(M_\odot)$ & $0.2$\\
$q_{\rm bin}=M_2/M_1$ & $0.29$\\                        
$\abin$ (AU)  & $0.22$  \\
$e_{\rm bin}$ & $0.16$  \\

\hline                                             
\end{tabular}
\end{table}

\subsubsection{Solid component}

In this work, dust is modelled as a second fluid whose governing equations are given by:
 \begin{equation}
 \frac{\partial \rho_d}{\partial t}+{\bf \nabla}\cdot(\rho_d{\bf v_d})=0,
 \end{equation}
 \begin{equation}
 \rho_d\left(\frac{\partial {\bf v_d}}{\partial t}+{\bf v_d}\cdot{\bf \nabla} {\bf v_d}\right)=-\rho_d {\bf \nabla} \Phi+\rho_d{\bf f_{d}},
 \end{equation}
where $\rho_d$ is the density of solids and ${\bf v_d}$ is the velocity. 

\begin{figure}
\centering
\includegraphics[width=\columnwidth]{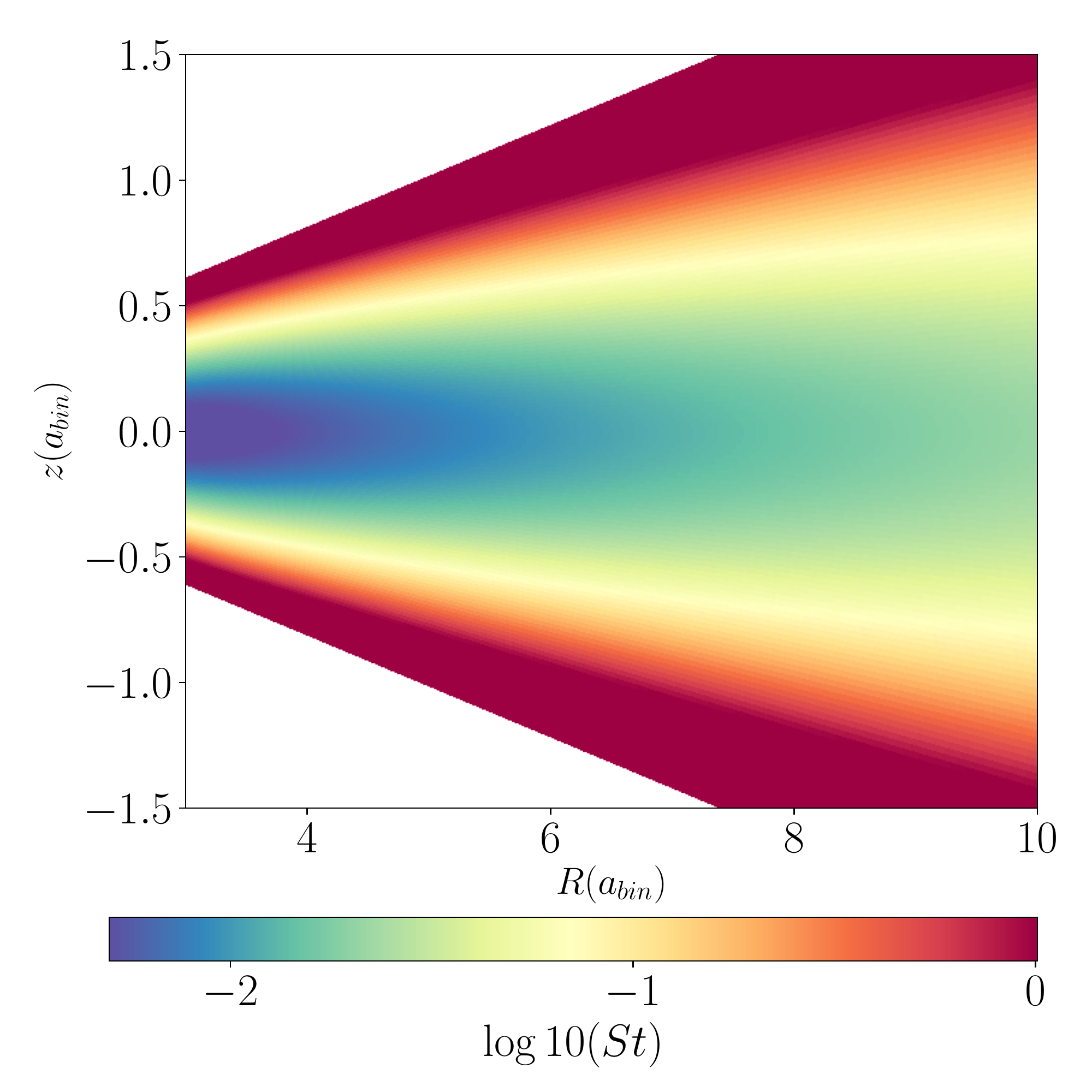}
\caption{Two-dimensional initial R-Z distribution of  the Stokes number of cm-sized particles for the initial disc model.}
\label{fig:stokes}
\end{figure}

In paper I, we considered particles that were characterized by a constant value of the Stokes number. In order to make physically meaningful predictions, here we fix  the particle  size  to a value $a=1cm$. For this particle size and the initial disc model, the corresponding Stokes number is shown in the right panel of Fig. \ref{fig:stokes}.  Particles experience gas drag according to the classical formulae for the Epstein and Stokes regimes. In the Epstein regime, namely for grains such that $\lambda>4a /9$ where $\lambda$ is the mean free path of the gas,  the drag force is given by: 
\begin{equation}
{\bf f_{d}}=\frac{1}{t_s}({\bf v}-{\bf v_d})
\end{equation}
with the stopping time
\begin{equation}
t_s=\frac{\rho_p a}{\rho c_s},
\end{equation}
where $c_s$ is the sound speed, $\rho_p$ is the material density, which is chosen to be $\rho_p=2g.cm^{-3}$. In the Stokes regime ($\lambda \le 4a /9$), the drag force is given by: 
\begin{equation}
{\bf f_{d}}=\frac{1}{2}C_D \pi a^2 \rho |{\bf v}-{\bf v_d}| ({\bf v}-{\bf v_d}),
\end{equation}
where $C_D$ is the drag coefficient, which is given by:
\begin{equation}
C_D=
\begin{cases}
24 {\cal R}_e^{-1}  & {\cal R}_e<1 \\
24 {\cal R}_e^{-0.6}  & {\cal R}_e<1\le 800 \\
0.44,
\end{cases}
\end{equation}	
where ${\cal R}_e$ is the Reynolds number defined by:
\begin{equation}
{\cal R}_e=\frac{2a|{\bf v_d}-{\bf v}|}{\nu_p},
\end{equation}
where $\nu_p=\lambda c_s/3$ is the molecular viscosity. The mean free path of the gas $\lambda=1/n\sigma_{H_2}$, with $n=\rho/(\mu m_H)$ the number density of particles and $\sigma_{H_2}=10^{-15}cm^2$ the collisional cross-section of molecular hydrogen , is computed assuming a mean molecular weight of $\mu=2.4$.

\subsection{Numerical setup}
The simulations presented in this paper were performed using the multifluid version of FARGO3D (Benitez-Lamblay \& Masset 2016). 
Computational units are chosen such that the total mass of the binary is $M_\star=1$, the gravitational constant $G=1$, and the radius $R=1$ in the computational domain corresponds to the binary semi-major axis for the  Kepler-16 system ($\abin= 0.22$ au, see Table 1). When presenting the simulation results, unless otherwise stated we use the binary orbital period $\Tbin=2\pi\sqrt{\abin^3/GM_\star}$ as the unit of time.

 The computational domain in the radial direction extends from $R_{\rm in}=1.5\;\abin$ to $R_{\rm out}=11\;\abin$ and we employ $716$ logarithmically spaced grid cells. In the azimuthal direction the simulation domain extends from $0$ to $2\pi$ with $700$ uniformly spaced grid cells. In the meridional direction, the simulation domain covers $3.5$ disc pressure scale heights above and below the disc midplane, and we adopt $144$ uniformly spaced grid cells.  Unstable inertial modes that are involved in the parametric instability have vertical wavelength $\lambda_z\sim H$,  and radial wavelength $\lambda_R\sim 0.6 H$ (Papaloizou 2005) and these  are resolved by $\sim 18$ grid cells in the vertical direction and $\sim 11$ grid cells in the radial one. The numerical resolution we adopt in this work is therefore similar to that employed in Paper I.

\subsection{Initial conditions}
\label{sec:init}

The initial radial profile of the sound speed, $\cs$, is given by: 
\begin{equation}
\cs(R)=h_0\left(\frac{R}{R_0}\right)^{q/2},
\end{equation}
where $R=r\sin \theta$ is the cylindrical radius and $h_0$ the disc aspect ratio at $R=R_0=1$. We adopt $h_0=0.05$ and $q=-1$ such that the aspect ratio, $h$, is constant with $h=h_0=0.05$.

The initial density and azimuthal velocity profiles of the gas are given by:
\begin{equation}
\rho(R,z)=f_{\rm gap}\, \rho_0\left(\frac{R}{R_0}\right)^p \exp\left(\frac{GM_\star}{\cs^2}\left[\frac{1}{\sqrt{R^2+z^2}}-\frac{1}{R}\right]\right)
\label{eq:rho0}
\end{equation}
and
\begin{equation}
v_\varphi(R,z)=\left[(1+q)\frac{GM_\star}{R}+(p+q)\cs^2-q\frac{GM_\star}{\sqrt{R^2+z^2}}\right]^{1/2},
\end{equation}
\label{sec:initial}
where $z=r\cos \theta$ is the altitude and $\rho_0$ the density at $R=R_0$. The power-law index for the density is set to $p=-5/2$ so that the slope of the surface density profile $\Sigma$ corresponds to that of the MMSN, namely $\Sigma \propto R^{-3/2}$ (Hayashi 1981). In Eq.~\ref{eq:rho0}, $\rho_0$ is also defined such that it corresponds to  the MMSN. $f_{\rm gap}$ is a gap-function used to initiate the disc with an inner cavity (assumed to be created by the binary), and is given by:
\begin{equation}
f_{\rm gap}=\left(1+\exp\left[-\frac{R-R_{\rm gap}}{0.1R_{\rm gap}}\right]\right)^{-1},
\end{equation}
 where $R_{\rm gap}=2.5\abin$ is the analytically estimated gap size (Artymowicz \& Lubow 1994). The initial radial and meridional velocities are set to zero. 
 
Regarding the dust component, the initial azimuthal velocity of the dust grains is Keplerian, and the initial radial and vertical  velocities are set to be zero.  The initial density of the grains is set  to $\rho_d(R,z)=\epsilon \rho(R,z)$, where $\epsilon$ is the dust-to-gas ratio. Initially, the dust-to-gas ratio therefore equals the metallicity $Z$ for which we consider initial values of $Z=0.01, 0.1$.

\subsection{Boundary conditions}
For both the gas and solid components, we employ a closed radial boundary condition at both the inner and outer edges of the disc. At the outer boundary, we also make use of a wave-killing zone for $R> 10$ to avoid wave reflection.  Ordinarily we would adopt outflow conditions at the inner edge to allow mass to accrete onto the binary, and hence for a steady state of the gas structure to develop. In these 3D calculations we consider inviscid conditions, such that a steady structure in which viscous stresses and gravitational torques come into balance does not exist. Furthermore, the computational expense of running 3D simulations would not allow us to achieve such a steady state even if we relaxed the inviscid assumption. We therefore do not expect the inner boundary condition to play an important role in determining the outcome of our simulations.

At the meridional boundaries, an outflow boundary condition is used for the velocities, dust density, and gas internal energy, and for which all  quantities in the ghost zones have the same values as in the first active zone, except the meridional velocity whose value is set to $0$ if it is directed towards the disc midplane to prevent inflow of material. For the  gas density, we follow Bae et al. (2016a) and maintain vertical stratification by solving the following condition for hydrostatic equilibrium in the meridional direction:
\begin{equation}
\frac{1}{\rho}\frac{\partial}{\partial \theta} (\cs^2 \rho)=\frac{v_\phi^2}{\tan \theta}.
\end{equation}

\subsection{Diagnostics}

To analyse the results of the simulations, we define at each altitude $z=r\cos \theta$ the mean value $<Q>_z$ of a quantity $Q$ by space averaging $Q$ over the full azimuth $2\pi$ and over a small radial range around centred on a reference radius $R_0$:
\begin{equation}
<Q>_z=\frac{\int_0^{2\pi}\int_{r_{in}}^{r_{out}}Qr \sin \theta dr d\phi}{\int_0^{2\pi}\int_{r_{in}}^{r_{out}}r \sin \theta dr d\phi}
\end{equation}
where $r_{in}$ and $r_{out}$ correspond to the integration boundaries. Similarly, we define at each radius $r$ a local value for $Q$ obtained by space averaging $Q$ over the full azimuth $2\pi$ and meridional direction $\theta \in [\theta_{min}, \theta_{max}]$:
\begin{equation}
<Q>_r=\frac{\int_0^{2\pi}\int_{\theta_{min}}^{\theta_{max}}Q \sin \theta d\theta d\phi}{\int_0^{2\pi}\int_{r_{in}}^{r_{out}} \sin \theta d\theta d\phi}
\end{equation}

 We also denote by $\overline{<Q>_i}$ the temporal average of the quantity $<Q>_i$  over time:
 \begin{equation}
\overline{<Q>_i}=\frac{1}{t_2-t_1}\int_{t_1}^{t_2}  <Q>_i dt
\end{equation}
where $t_1$ and $t_2$ set the limits of integration in time.

\section{Results}
\subsection{Hydrodynamical turbulence driven by the disc eccentricity and the impact of the dust back-reaction}
\label{sec:turbulence}

\begin{figure}
\centering
\includegraphics[width=\columnwidth]{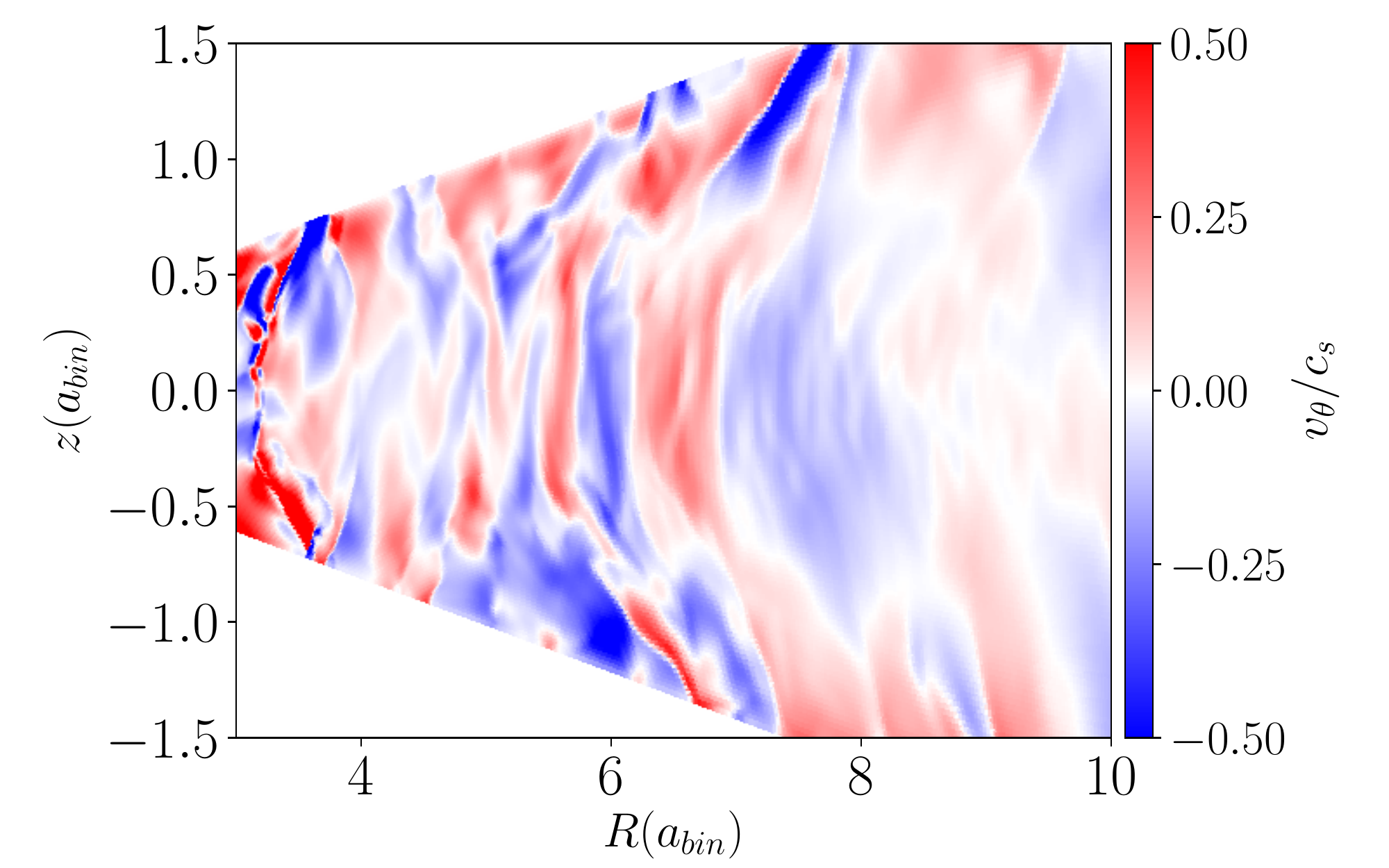}
\caption{Contours of the gas meridional velocity $v_\theta$ in the $[R,Z]$ plane in the case where dust feedback is discarded}
\label{fig:vtheta}
\end{figure}

\begin{figure}
\centering
\includegraphics[width=\columnwidth]{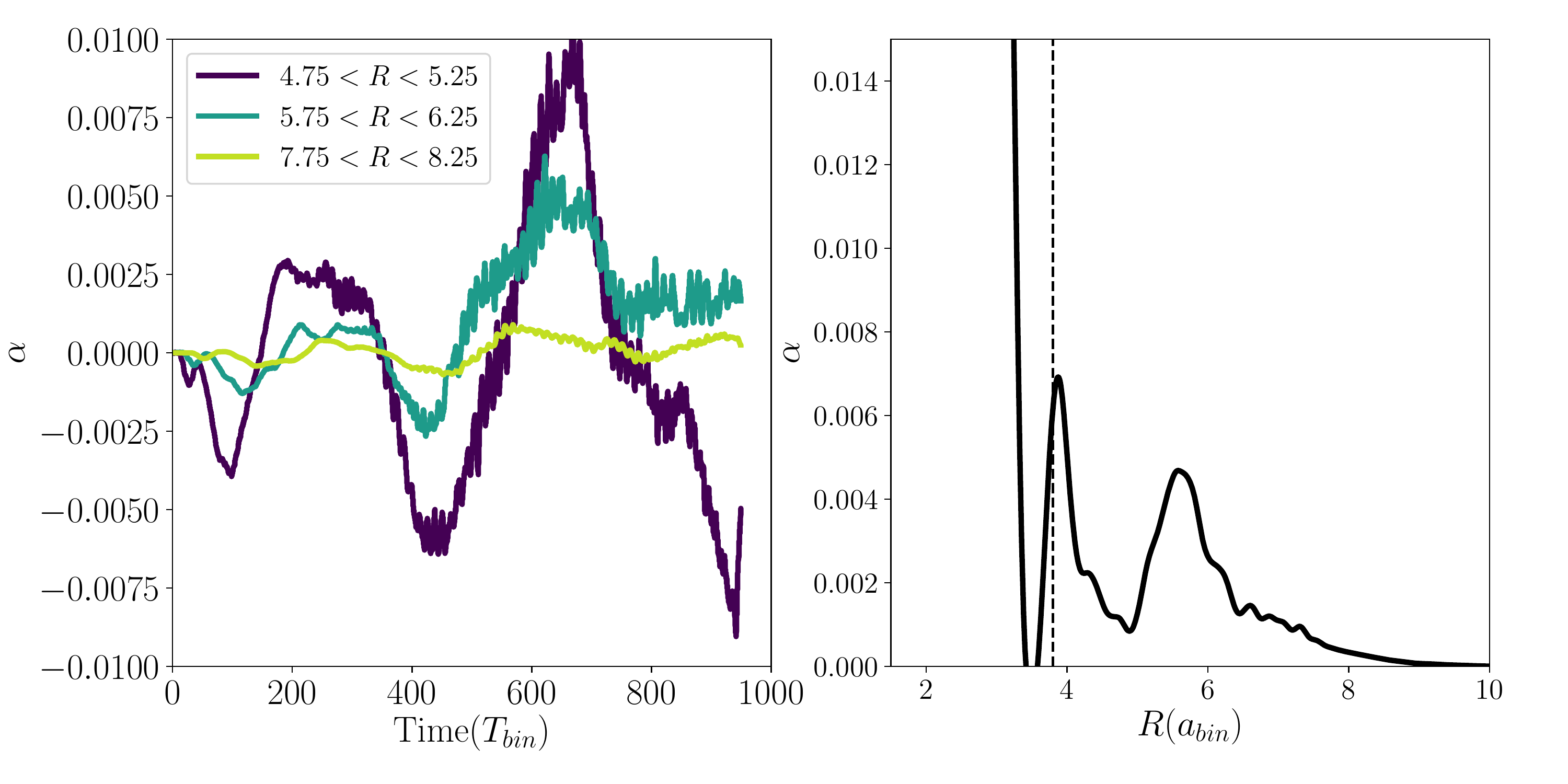}
\includegraphics[width=\columnwidth]{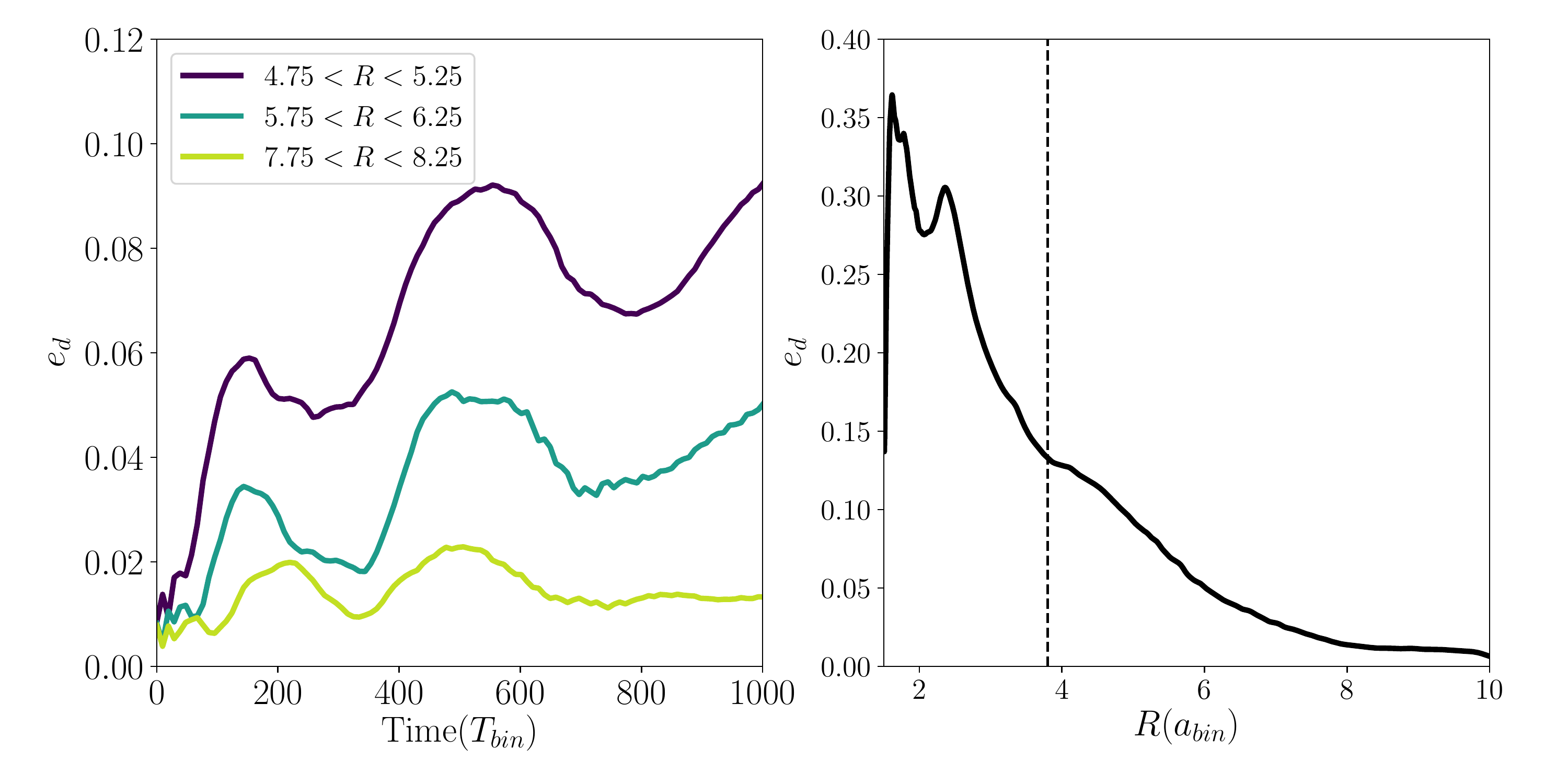}
\caption{{\it Left}: $\alpha$ parameter (top) and disc eccentricity (bottom) as a function of time for the model without back-reaction included. {\it Right}: Radial profile of $\alpha$ (top) and $e_d$ (bottom) in the disc midplane. Here,   a temporal average of 400 binary orbits has been used. The vertical dashed line corresponds to the location of the edge of the inner cavity.}
\label{fig:alpha_ecc}
\end{figure}

Eccentric discs are unstable to a parametric instability  involving the resonant interaction between inertial-gravity waves and the eccentric mode in the disc (Papaloizou 2005a; Barker \& Ogilvie 2014). The instability occurs whenever the inertial wave frequency $\omega$ matches the resonance condition:

\begin{equation}
\omega \sim \frac{\Omega}{2}
\end{equation}

Both global (Papaloizou 2005b) and local (Wienkers \& Ogilvie 2018) simulations have found that the non-linear outcome of the instability is turbulence, with corresponding $\alpha$ viscous stress parameter that depends on the value for the disc eccentricity $e_d$. For instance, Papaloizou (2005) found $\alpha\sim 10^{-3}$  for $e_d=0.1$, whereas for smaller disc eccentricities $e_d=0.03$, $\alpha\sim 2\times 10^{-4}$ (Wienkers \& Ogilvie 2018).
In circumbinary discs where an eccentricity mode in the disc can grow as a result of the interaction with the central binary, our previous work demonstrated that this instability can also operate and ultimately lead to turbulence. For a central binary with parameters typical of Kepler-16 and constant cooling timescale $t_{cool}=\Omega^{-1}$, we confirmed the value for $\alpha$ to be correlated with the disc eccentricity, with $\alpha \sim 5\times 10^{-3}$ at maximum.

For the  more realistic thermal disc structure we consider here, turbulence is also found to set in, and with properties similar to what has been found in Paper I. In particular, the vertically elongated flow structures that were obtained in Paper I are reproduced, as revealed by looking  at contours of the meridional velocity in  Fig. \ref{fig:vtheta}. These vertical gas motions have typical radial wavelength of $\lambda_R \sim H$, consistent with what is expected from linear theory (Papaloizou 2005). 

In the case where the effect of dust back-reaction is neglected, the joint temporal evolution of $\alpha$ and $e_d$   is shown, for different radial bins of the disc in the left panel of Fig. \ref{fig:alpha_ecc}.  To compute $\alpha$, we first average the Reynolds stress  $T_{r\phi}=\rho  \delta v_r \delta v_\phi$ over the azimuthal and meridional directions, where $\delta v_r$ and $\delta v_\phi$ are the radial and azimuthal velocity fluctuations. A local value for $\alpha$ is then given by $\alpha(r)=<T_{r \phi}>_r/<P>_r$ where $<P>_r$ is a mean pressure.  Both $\alpha$ and $e_d$  undergo oscillations around a mean value because the binary longitude of pericentre maintains a constant value while the disc tends to precess with a finite libration amplitude. A  clear trend for $\alpha$ to increase with disc eccentricity can be noticed, as expected for turbulence originating from the eccentricity induced parametric instability. This can also be observed in the right panel of Fig.  \ref{fig:alpha_ecc} which depicts the azimutally averaged radial profiles of both quantities at the disc midplane, and further time-averaged over one precession period, which is estimated to $\sim 400$ $T_{bin}$.  Near the cavity edge where particles are expected to concentrate, $\alpha \approx 0.01$ at maximum, which corresponds to a significant level of turbulence. At larger distances, however, where the disc eccentricity is much smaller than at the cavity edge, we see the turbulence drops off significantly such that particle stirring should also decrease there.

Although we will examine in detail the vertical distribution of particles later in the paper, we can anticipate that cm-sized pebbles with $\st\approx 0.01$ located close to the cavity edge should experience significant stirring, such that solid particles would have more or less same scale height as the gas. The main consequence is that this would render pebble accretion inefficient, since the pebble accretion rate scales as $r_H/H_d$, where $r_H$ is the Hill radius. In particular, for a Ceres-mass body of mass $\sim 10^{-4}M_\oplus$, a simple estimate suggests that the accretion rate would drop by a factor of $\sim 100$ compared to when the solid particles are more settled.
    
\begin{figure}
\centering
\includegraphics[width=\columnwidth]{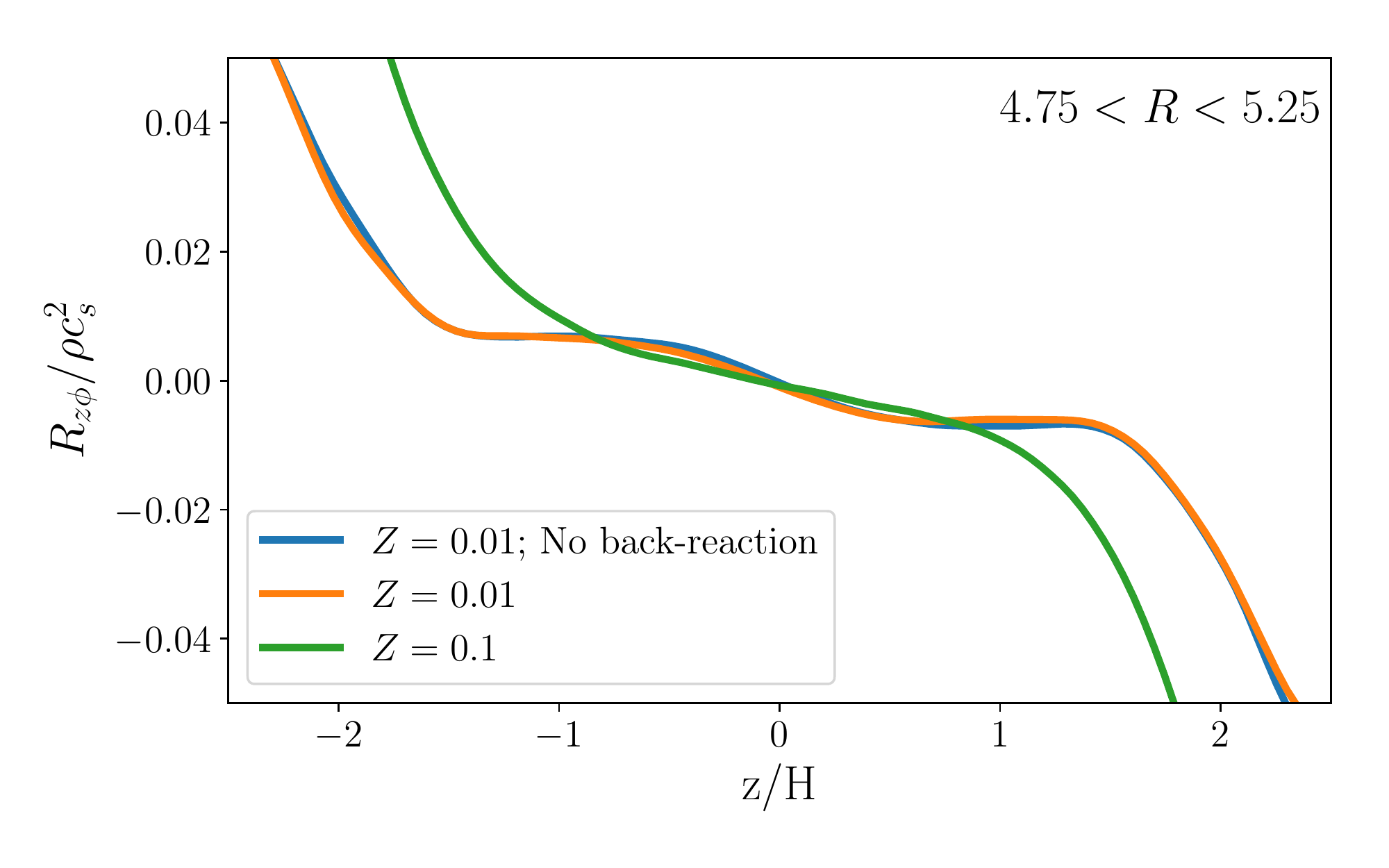}
\includegraphics[width=\columnwidth]{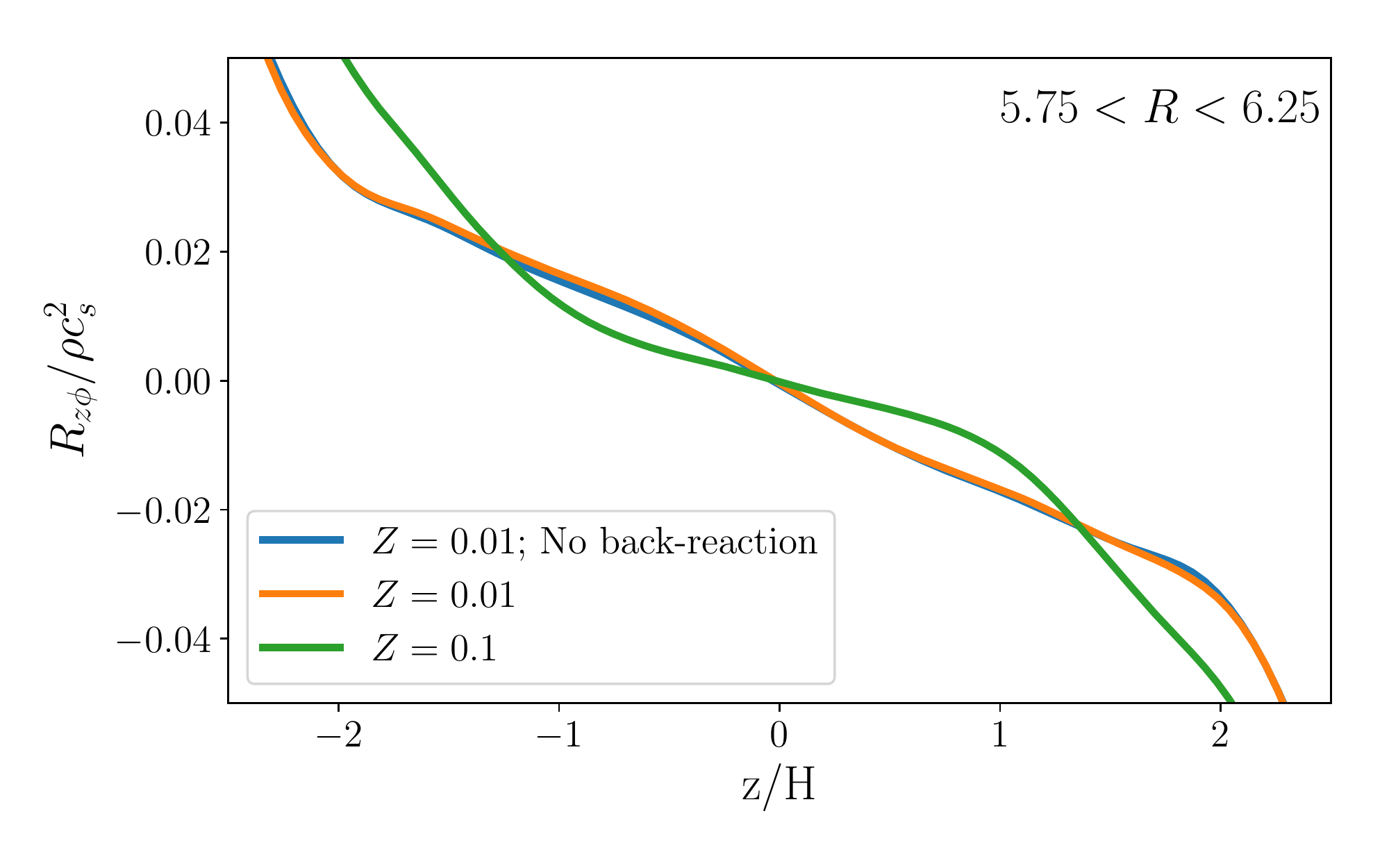}
\caption{{\it Top:} Vertical profile of the vertical stress $R_{z\phi}$ in the region $R\in[4.75,5.25]$ for the models with and without the effect of the dust back-reaction onto the gas included. {\it Bottom:} same but for $R\in[5.75,6.25]$. }
\label{fig:Rzphi}
\end{figure}

To assess the effect of the dust back-reaction on the eccentricity-induced parametric instability,  we present in Fig~\ref{fig:Rzphi} the mean vertical profile of the vertical stress:
\begin{equation}
R_{z\phi}=\overline{<\rho \delta v_\phi \delta v_z>_r}
\end{equation}
for our various models with/without back-reaction, where $\delta v_z$ is gas vertical velocity fluctuation. For anisotropic turbulence dominated by large scale vertical motions, as is the case here,  the vertical stress is more relevant to the classical radial stress to quantity  turbulence strength (Stoll \& Kley 2017; Lin 2019).  Here, two different regions of the disc is considered for  the radial average  and a temporal average over $400$ binary orbits is also used.  It is clear that the primary effect of the particle back-reaction is to stabilize the instability, as a higher dust abundance results in a smaller vertical stress near the disc midplane.  Compared to  regions of the disc located at $R\approx 6$, the effect of the particle back-reaction and metallicity are slightly weaker at $R\approx 5$ because the dust scale height tends to be higher there (see Sect. \ref{sec:vertical}). Hence, we can conjecture that a higher metallicity would tend to induce a smaller level of turbulence and therefore to favour dust settling, resulting in an enhanced efficiency of pebble accretion. As noted earlier, a similar result was found by Lin (2019) in the case of turbulence driven by the VSI.

\begin{table*}
\caption{Properties of the hydrodynamical turbulence as a function of parameters in the region $5.75<R<6.25$}  
\centering                                        
\begin{tabular}{c c c c c c c c c}          
\hline\hline                        
  Dust-to-gas ratio & Back-reaction & $H_d/H$ & $<\overline{\delta v_{g,z}}>$   & $\alphagz=<\overline{\deltavgz}>\tau_c$ &  $H_d/H$ & $\ddz/c_sH$  & $H_d/H$ \\ 
                           &                             &                &   $(c_s)$ & & (Eq. \ref{eq:hd})& (Eq. \ref{eq:ddz}) & \\
\hline 
$Z=0.01$ & No & $0.52$ & $0.19$ & $0.018$ & $0.8$ & $0.01$ &$0.7$ \\
$Z=0.01$ & Yes & $0.52$ &$0.19$ & $0.018$ & $0.8$  & $0.01$& $0.7$\\
$Z=0.1$ & Yes & $0.43$ &$0.14$ & $0.01$ & $0.7$  & $0.005$ &$0.6$\\                    
\hline                                             
\end{tabular}
\label{table2}
\end{table*}

\subsection{Vertical distribution of cm-sized particles}
\label{sec:vertical}

\begin{figure}
\centering
\includegraphics[width=\columnwidth]{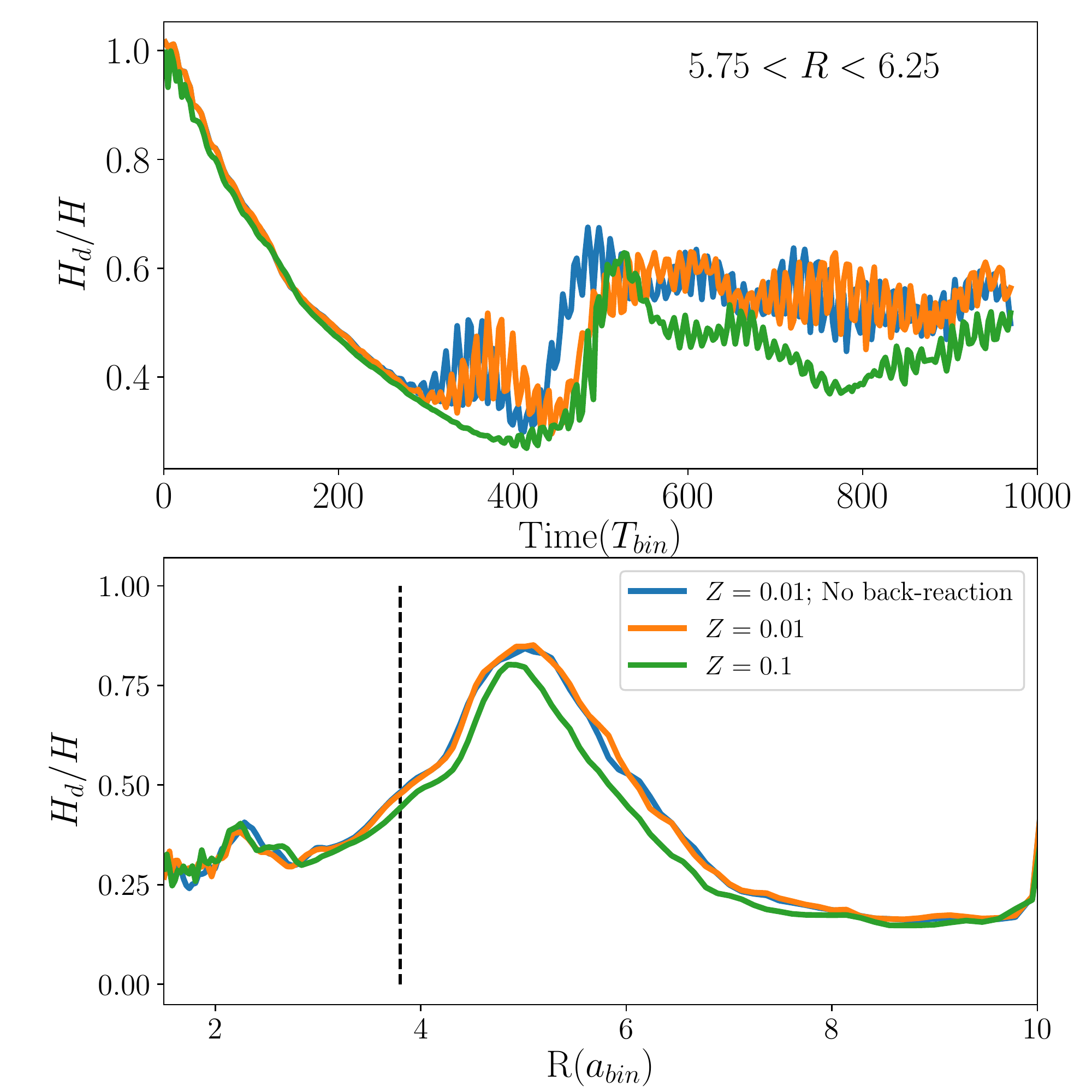}
\caption{{\it Top:} Temporal evolution of the dust scale height relative to the the gas scale height. {\it Bottom: }  Dust scale height as a function of radius. The vertical dashed line corresponds to the location of the edge of the inner cavity.}
\label{fig:hd}
\end{figure}

\begin{figure*}
\centering
\includegraphics[width=0.33\textwidth]{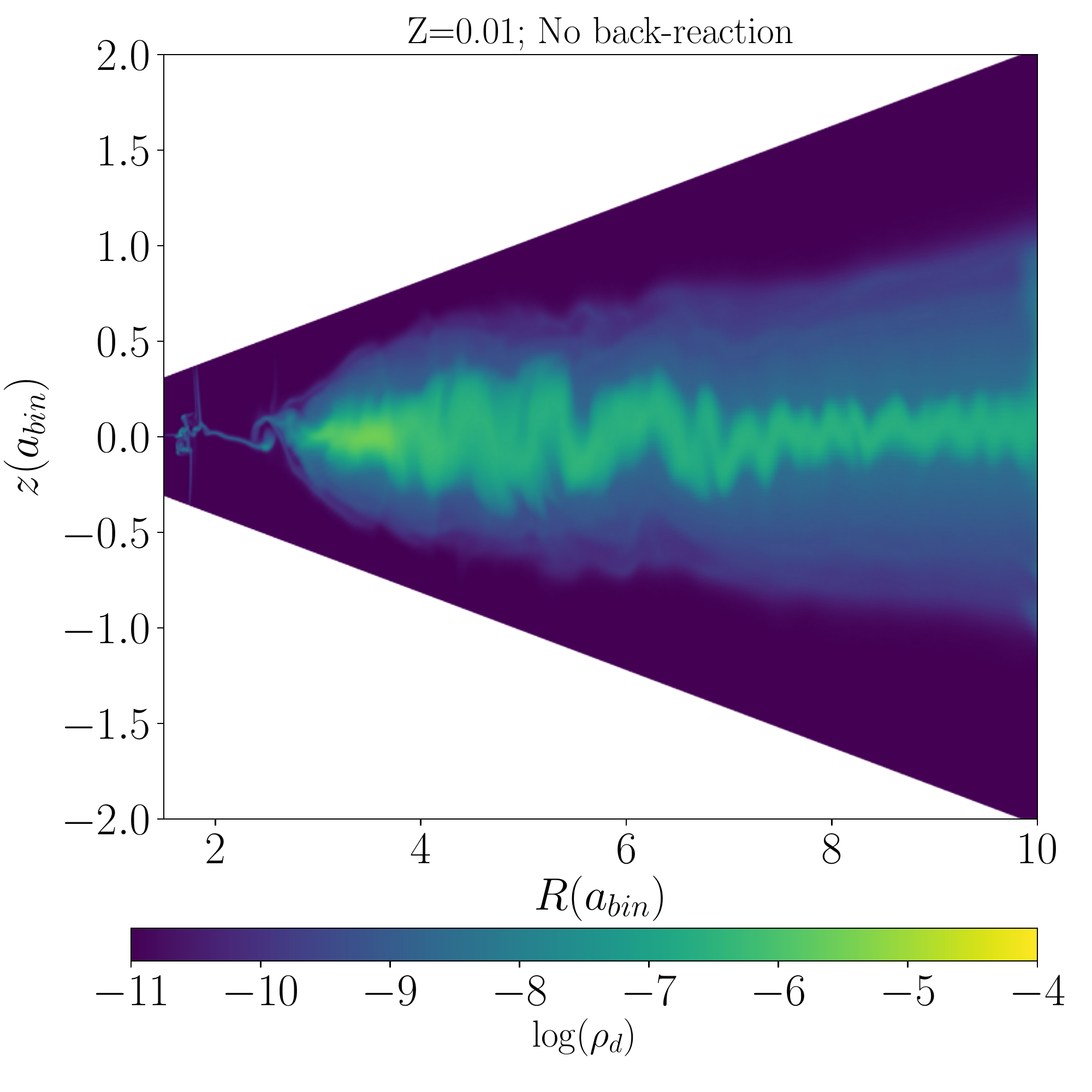}
\includegraphics[width=0.33\textwidth]{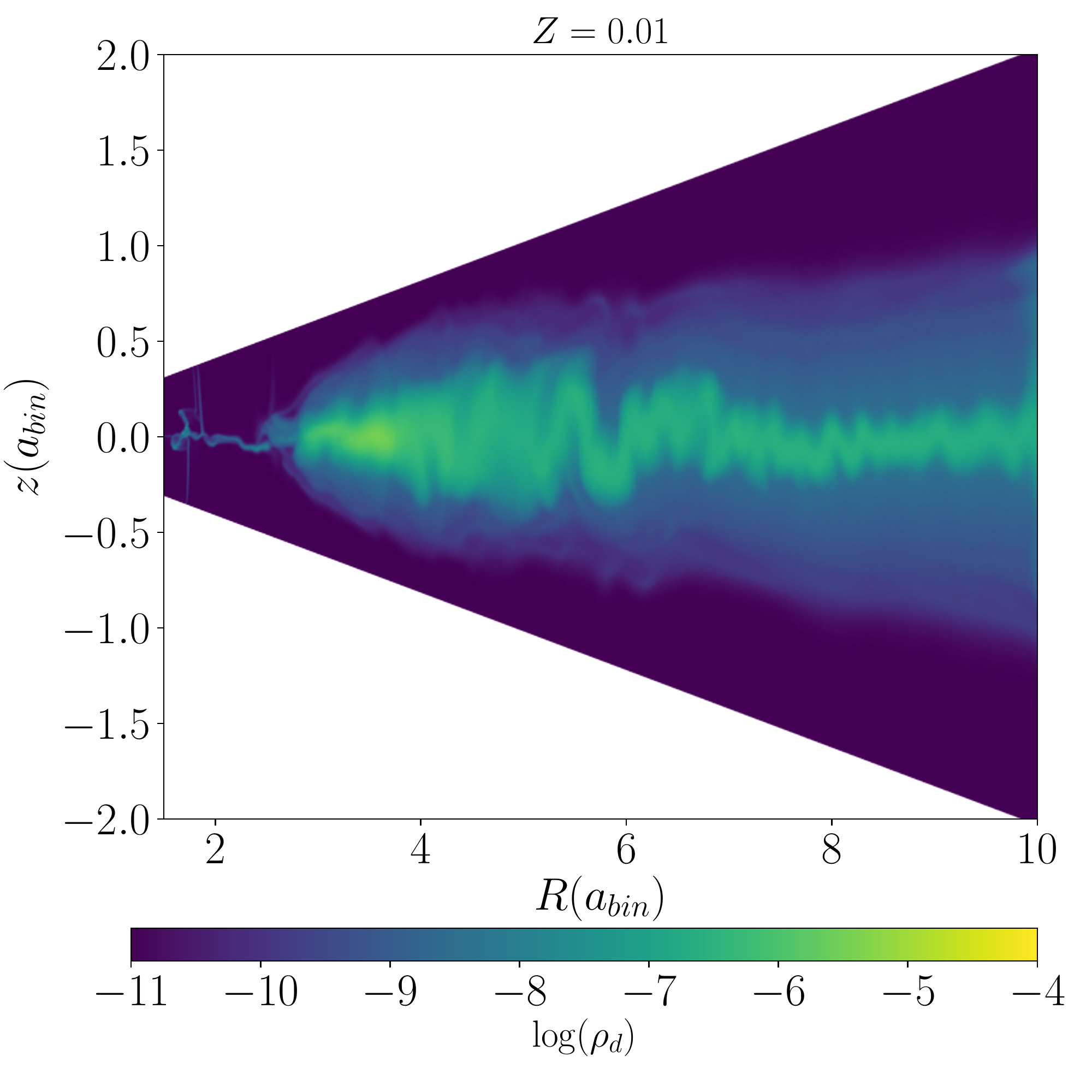}
\includegraphics[width=0.33\textwidth]{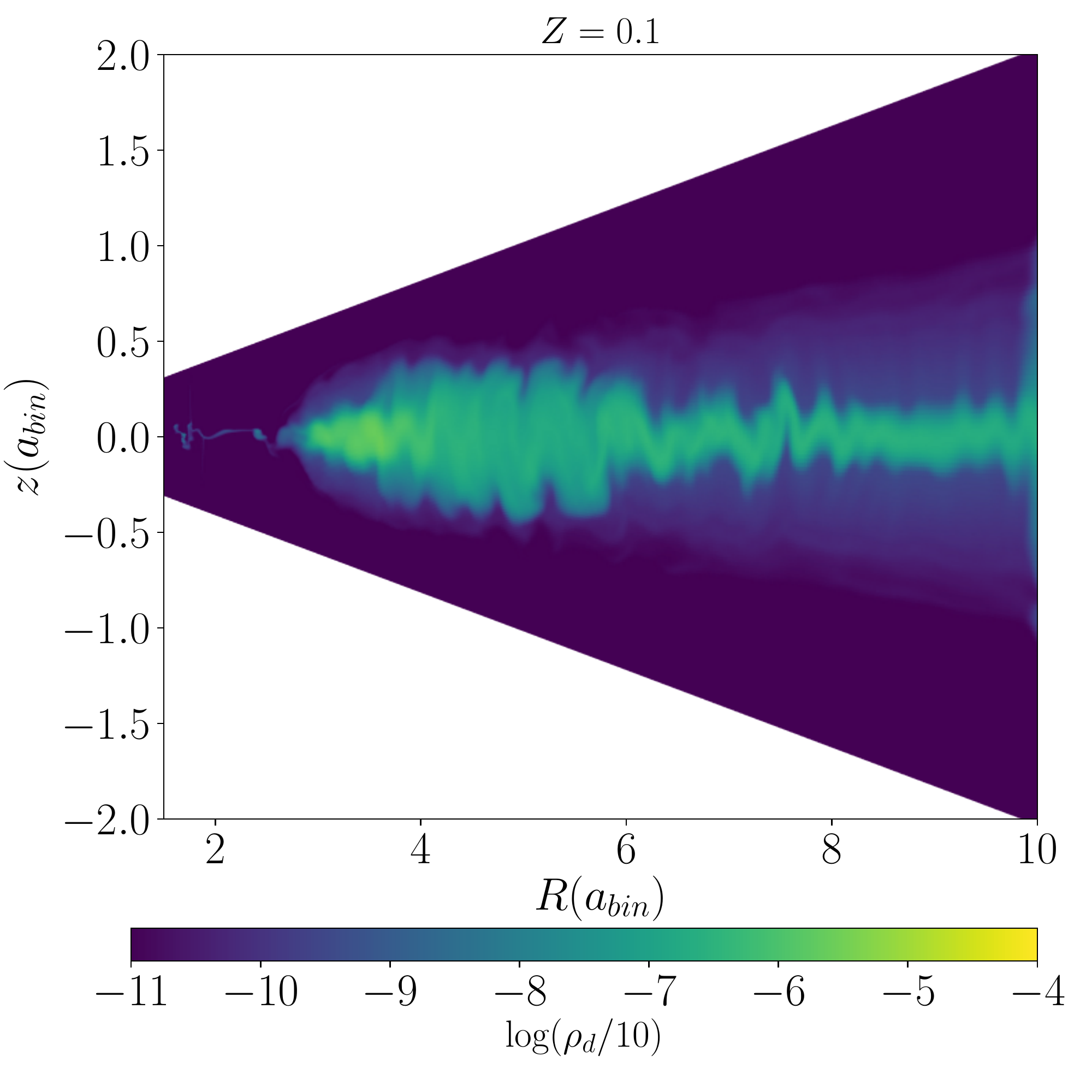}
\includegraphics[width=0.33\textwidth]{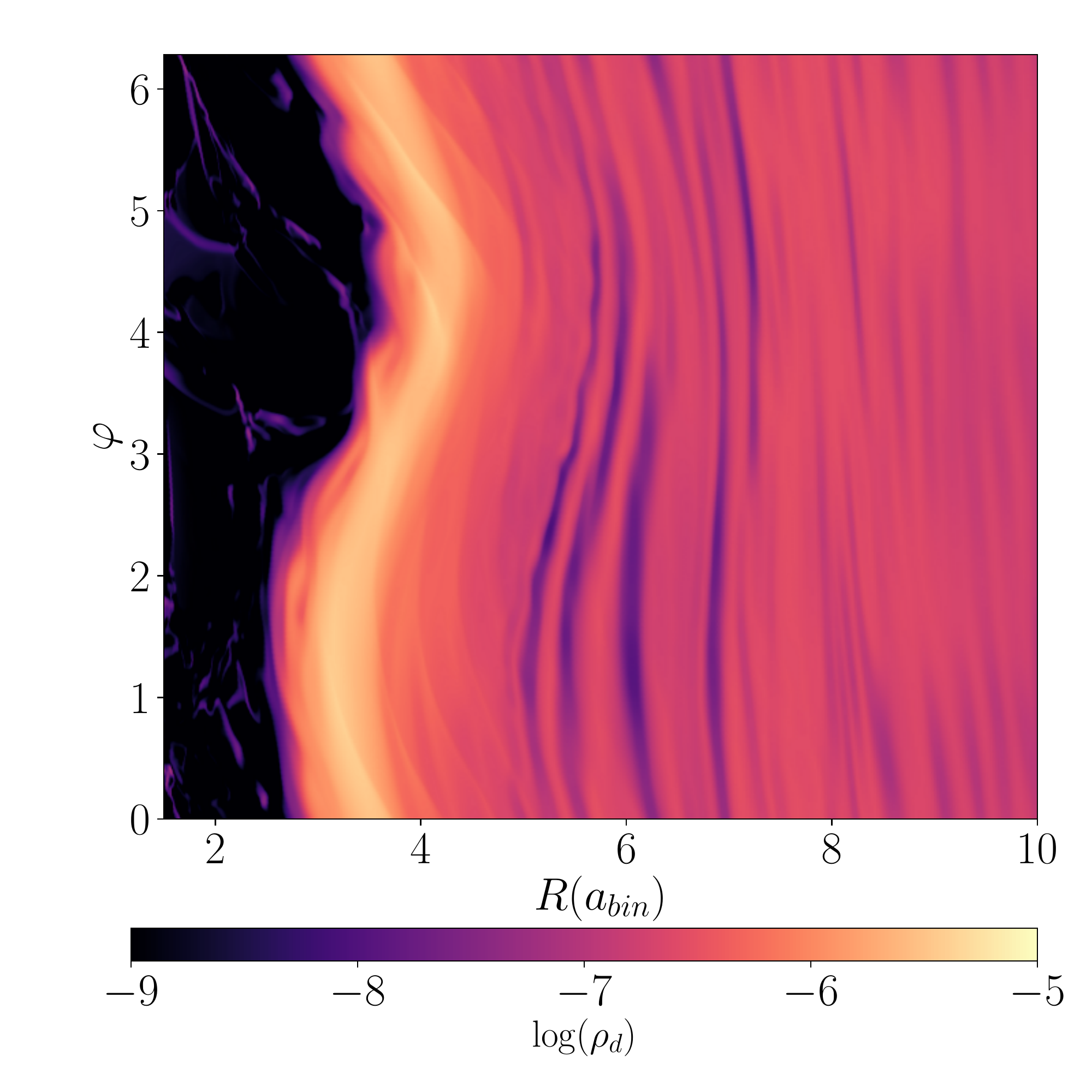}
\includegraphics[width=0.33\textwidth]{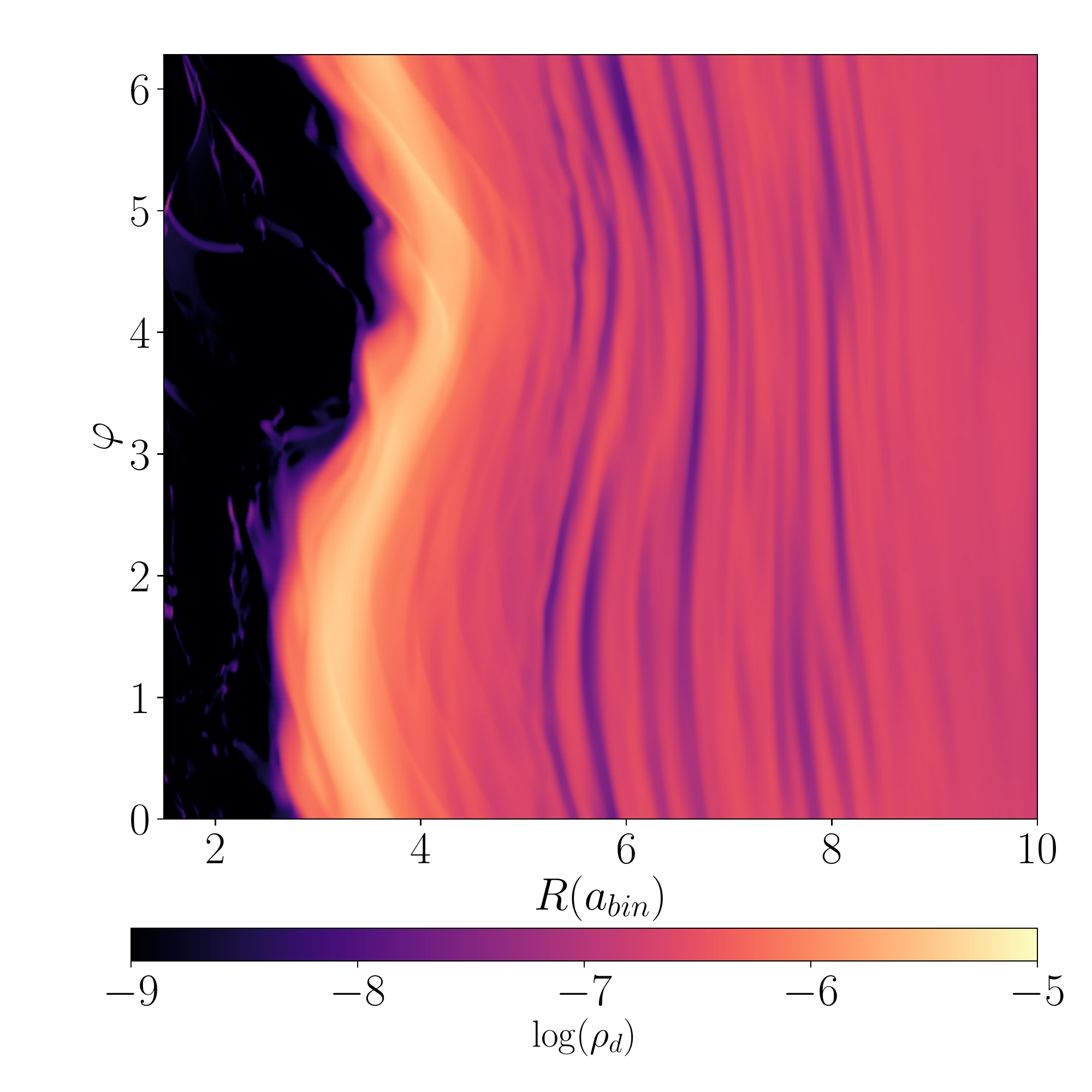}
\includegraphics[width=0.33\textwidth]{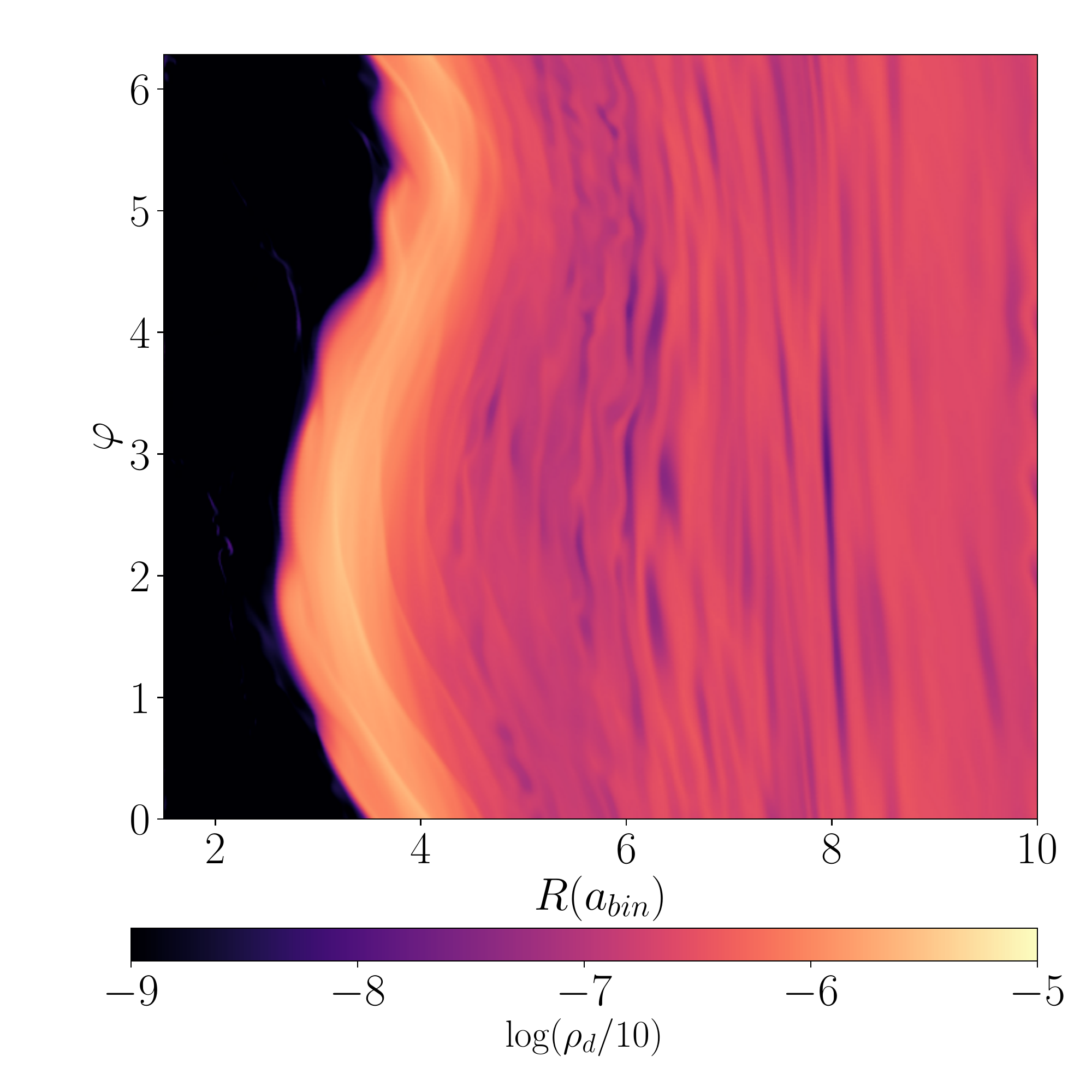}
\caption{{\it Top:} vertical distribution of dust particles at $t=1000$ for the three models we considered. {\it Bottom:} density of solids in the disc midplane for the three models.  In the case with $Z=0.1$, the dust density has been scaled by a factor of $0.1$ for easier comparison.}
\label{fig:2dplot}
\end{figure*}

In order to examine the effect of the particle back-reaction in more detail, we now focus on the dynamical evolution of embedded cm-sized particles. We plot  in the upper panel of Fig. \ref{fig:hd} the temporal evolution of the dust scale height averaged between $R=5.75$ and $R=6.25$, relative to the gas scale height. The dust scale height  $H_d$ is determined by fitting the dust distribution by a Gaussian function, which from Paper I appears to be a fairly good approximation.  As expected, particles initially settle to the disc midplane until the onset of turbulence prevents further settling. From this point in time,  the vertical distribution of particles reaches a steady state,  as gravitational settling tends to balance turbulent diffusion. The vertical distribution of solids at the end of each run is shown in the top panel of Fig. \ref{fig:2dplot}, while the particle distribution at the disc midplane is presented in the bottom panel. As the particles that we consider here are marginally coupled to the gas, they become radially concentrated at the edge of the inner cavity and can be concentrated by the spiral waves launched by the binary.

Figure~\ref{fig:hd} shows that consistent with the above discussion about the effect of the solid abundance on the vertical stress, we obtain a smaller dust scale height for the run with higher metallicity $Z=0.1$.  At equilibrium, we find that the difference in metallicity in the interval $R\in[5.75,6.25]$ is  a factor of $\sim 12$ higher between the cases $Z=0.01$ and $Z=0.1$, which confirms that dust settling is favored in this latter case. However, the effect of the particle back-reaction appears to be negligible for a smaller metallicity $Z=0.01$. For each model, the values for $H_d/H$ at steady-state are listed in Table \ref{table2}.  

\subsection{Comparison with analytical estimates}
\label{sec:analytical}
An issue that is worth examining is how the ratio $H_d/H$ obtained in our simulations at steady state compares with various analytical estimates that have appeared in the literature. An estimate obtained using diffusion theory (Dubrulle et al. 1995; Youdin \&  Lithwick 2007) is given by:
\begin{equation}
\frac{H_d}{H}=\sqrt{\frac{\alphagz}{\alphagz+\st}}
\label{eq:hd}
\end{equation}
where $\alphagz=\dgz/\Omega H^2$ is the dimensionless vertical diffusion coefficient of the gas.  The gas vertical diffusion coefficient $\dgz$ is given by: 
\begin{equation}
D_{g,z}\sim <\overline {\delta v_{z,g}^2}> \tau_{cor},
\label{eq:dgz}
\end{equation}
where $ \delta v_{z,g}$ is the vertical velocity dispersion of the gas and $\tau_{cor}$ the  correlation timescale of the hydrodynamical turbulence.  $\tau_{cor}$ can be determined by computing the time integral of the autocorrelation function (ACF) of the meridional velocity  (Yang et al. 2009). This procedure has been used in Paper I and leads to $\tau_{cor}\sim0.015$  $T_{orb}$. 

\begin{figure}
\centering
\includegraphics[width=\columnwidth]{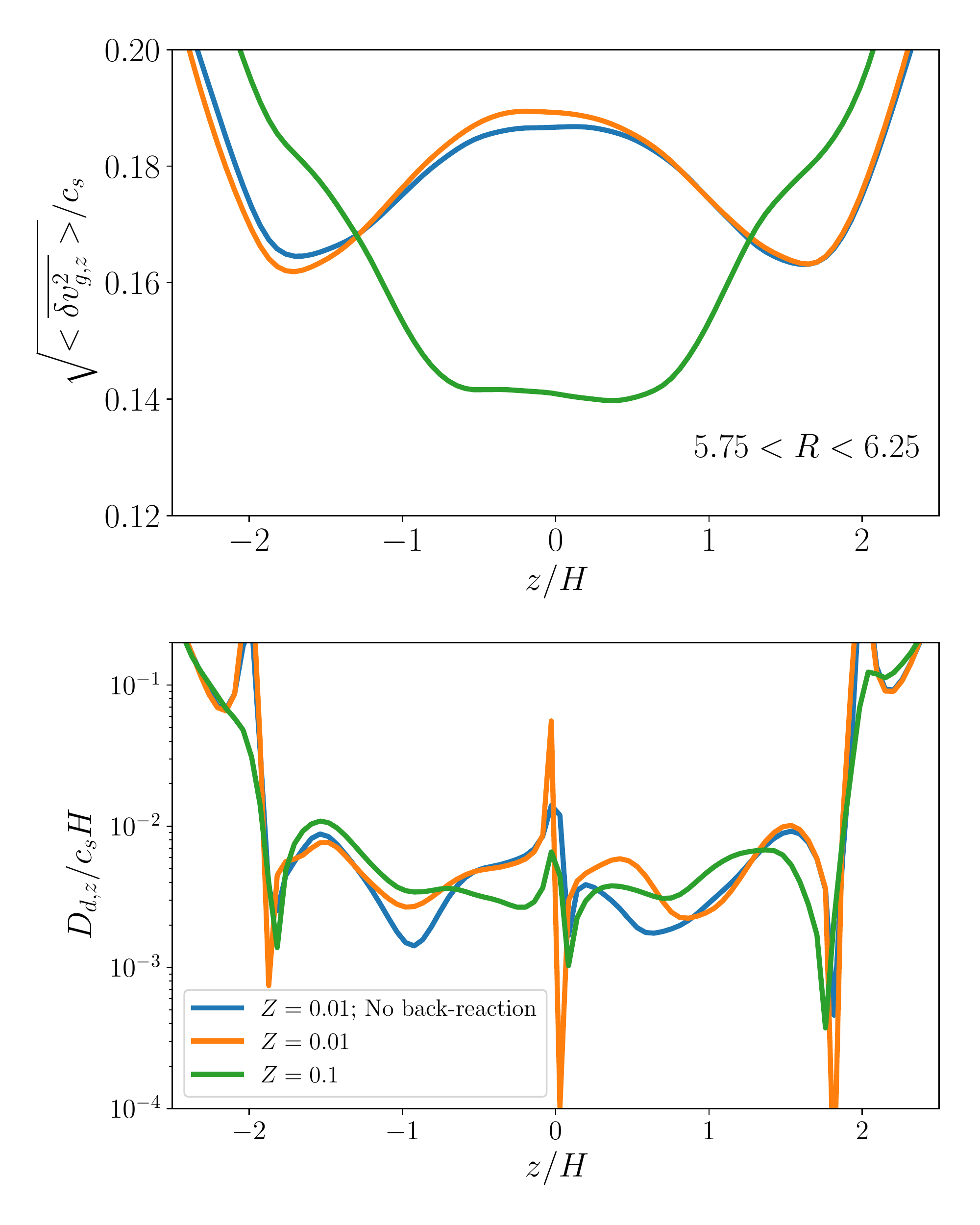}
\caption{{\it Top:} mean gas vertical velocity fluctuations as a function of height for the three models we considered.  {\it Bottom:} vertical profile of the dust diffusion coefficient computed using Eq. \ref{eq:ddz}}
\label{fig:dcoef_tau}
\end{figure}

We plot the mean vertical velocity dispersion as a function of $z$ in the upper panel of Fig. \ref{fig:dcoef_tau}.  We immediately remark that  this quantity: i) tends to increase (resp. decrease) when moving away from the disc midplane for high (resp. low) metallicities; ii) has a smaller value in the disc midplane for higher dust-to-gas ratios. For each model, the values for $\mdeltavgz$  at the disc midplane are reported in the fourth column of Table~\ref{table2}, while the resulting values for $\alphagz$ are listed in the fifth column. Given that the cm-sized particles located at $R\sim 6$ have $\st\sim 0.01$ (see Fig.\ref{fig:stokes}), plugging these latter values for $\alphagz$ in Eq.~\ref{eq:hd} results in estimations for $H_d/H$ that are slightly overestimated (see Table \ref{table2}).  Moreover, using  Eq. ~\ref{eq:hd} would predict a minimum in the dust scale height profile at $R\approx 5$, as viscous stresses tend to be smaller there (see top right panel of Fig. \ref{fig:alpha_ecc}). 

Such a discrepancy may be attributed to the fact that Eq.~\ref{eq:hd} assumes that the dust diffusion coefficient in the vertical direction $\ddz$ is equal to that of the gas,  which is reasonable as long as the solids are strongly coupled to the  gas. An alternative, more general estimation for $\ddz$ can be obtained by assuming that at steady-state, the mass flux due to vertical settling balances the mass flux due to turbulent diffusion (Dubrulle et al. 1995; Zhu et al. 2015), which leads to (Riols \& Lesur 2018):
\begin{equation}
D_{d,z}=-\frac{<\overline{\delta \rho_d \delta v_{z,d}}>}{(\overline \rho+\overline \rho_d) \langle\frac{\partial}{\partial z}\left(\frac{\overline \rho_d}{\overline \rho+\overline \rho_d}\right)\rangle}.
\label{eq:ddz}
\end{equation}
The vertical profile of $D_{d,z}$ is shown in the bottom panel of Fig.~\ref{fig:dcoef_tau},  and the values for $\ddz$ at the disc midplane are listed in  Table~\ref{table2}. Overall, using the expression for $\ddz$ given by Eq. \ref{eq:ddz} confirms the tendency for the dust diffusion coefficient to decrease as the dust-to-gas ratio is increased, although this leads to slightly smaller values in comparison to those given by Eq.~\ref{eq:dgz}. An alternative estimate for the dust scale height is then given by:
 \begin{equation}
\frac{H_d}{H}=\frac{1}{\sqrt{1+\frac{\st \Omega H^2}{D_{d,z}}}}.
\label{eq:hd2}
\end{equation}

\begin{figure}
\centering
\includegraphics[width=\columnwidth]{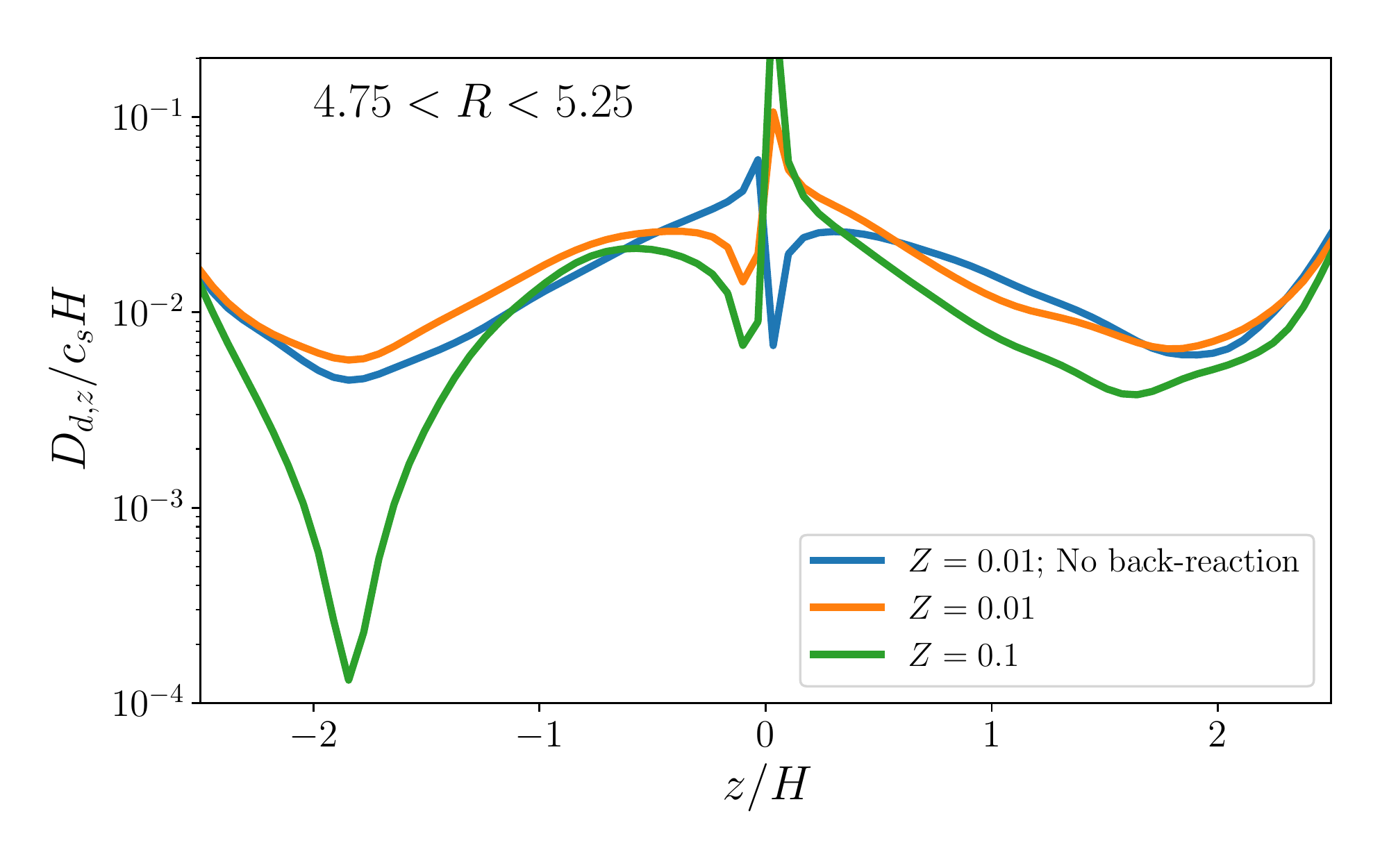}
\caption{Vertical profile of the dust diffusion coefficient computed using Eq. \ref{eq:ddz}, for $R\in[4.75,5.25]$.}
\label{fig:dcoef_tau_r5}
\end{figure}

Employing the previous equation results in a better agreement with the dust scale height deduced from the simulations (see Table \ref{table2}). For instance, for the model with $Z=0.1$ for which $H_d/H\approx 0.43$ at $R\approx 6$, using Eq. \ref{eq:hd} gives $H_d/H\approx 0.7$ whereas Eq. \ref{eq:hd2} leads to $H_d/H \approx 0.6$.   Moreover, looking at the vertical profile of $\ddz$ for $R\in[4.75,5.25]$  shown in Fig.~\ref{fig:dcoef_tau_r5} , we see that in this region  the dust diffusion is higher than further away in the disc, consistently with a higher dust scale height at this location. We note, however, that the estimates given by Eqs.~\ref{eq:hd} or \ref{eq:hd2}  are not expected to be completely accurate, as they assume a uniform dust diffusion coefficient, whereas Fig. \ref{fig:dcoef_tau} shows this is not the case. Moreover,  the diffusion model leading to these expressions assumes that the sizes of the turbulent eddies are smaller than $H$, while  the hydrodynamical turbulence operating here is rather a mixture of small scale turbulence and coherent vertical motions occurring on scales larger than $H$.

\subsection{Planet growth times}
\label{sec:GrowthTimes}
The smaller dust scale height obtained for higher dust-to-gas ratios confirms that the dust back-reaction tends to favour dust settling against the hydrodynamical turbulence. This is also illustrated by the bottom panel of Fig. \ref{fig:hd} which shows the radial profile of $H_d/H$ time averaged over $400$ $T_{bin}$.  At $R\approx 5$, namely close to the edge of the inner cavity,  a value of $H_d/H\sim 0.8$ for $Z=0.1$ is obtained whereas $H_d/H\sim 0.85$ for $Z=0.01$. This implies that a large dust abundance is required to significantly decrease the particle scale height. At this location, a planet whose Hill radius is larger than the dust scale height has planet-to-star mass ratio $q_p\sim 1.7\times 10^{-4}$  (or equivalently $\sim 50$ Earth masses)  and would experience efficient pebble accretion. In other words, this means that a planet with $q_p\lesssim 1.7\times 10^{-4}$ would accrete in the 3D regime for which the corresponding accretion rate, in the strong coupling limit,  is given by (Lambrechts et al. 2019):
\begin{equation}
\dot M_p\approx \frac{1}{4\sqrt{2\pi}}\frac{1}{\eta}\left(\frac{H_d}{H}\right)^{-1}\left(\frac{H}{r}\right)^{-1} F_{peb}q_p
\label{eq:mdot}
\end{equation}
where $F_{peb}$ is the pebble flux for which we adopt a nominal value of $F_{peb}=120M_\oplus/Myr$, and
\begin{equation}
\eta=-\frac{1}{2}h^2(p+q).
\end{equation}
Following Pierens et al. (2020), we can compute  the time, $t_{growth}$, required to grow a Ceres-mass planetesimal to a body with mass $m_{target}$ through pebble accretion. This is given by:
\begin{equation}
t_{growth}=\int_{m_{Ceres}}^{m_{target}}\frac{dm}{\dot M_p}.
\label{eq:tgrowth}
\end{equation}
Setting $m_{target}=10$~$M_\oplus$ results in $t_{growth}\sim 35$ $Myr$, much longer than expected disc lifetimes.   As mentionned earlier, this is because for  planet masses $\lesssim 50$ Earth masses, pebble accretion proceeds in the 3D regime. At $R\approx 8$, however,  $H_d/H\sim 0.25$ and the critical mass above which pebble accretion is not impacted by 3D effects becomes $q_p\sim 3\times 10^{-6}$. Using Eqs. \ref{eq:mdot} and \ref{eq:tgrowth}, we find that a core with $q_p\sim 3\times 10^{-6}$ can be formed in $\sim 6$ $Myr$ in that case. Above this value for $q_p$, the accretion of pebbles is expected to be very efficient as it proceeds in 2D, such that forming a 10 $M_\oplus$ body  within the disc lifetime seems very likely to be possible at large distances from the binary.

\subsection{Collision velocities}

Pebble accretion requires a pre-existing planetesimal seed which possibly formed through the streaming instability. The typical radius of planetesimals formed in models of streaming instability is 100 km (Schafer et al. 2017; Klahr \& Shreiber 2020). However, triggering the streaming instability at moderate solid-to-gas ratios requires the presence of large grains, which are not easily formed due to the bouncing and fragmentation barriers (Drazkowska \& Dullemond 2014).  As turbulence is a source of impact velocities, a question of interest is whether the collisions speeds induced by the turbulence remain low enough to allow for sticking and grain growth. 

\begin{figure}
\centering
\includegraphics[width=\columnwidth]{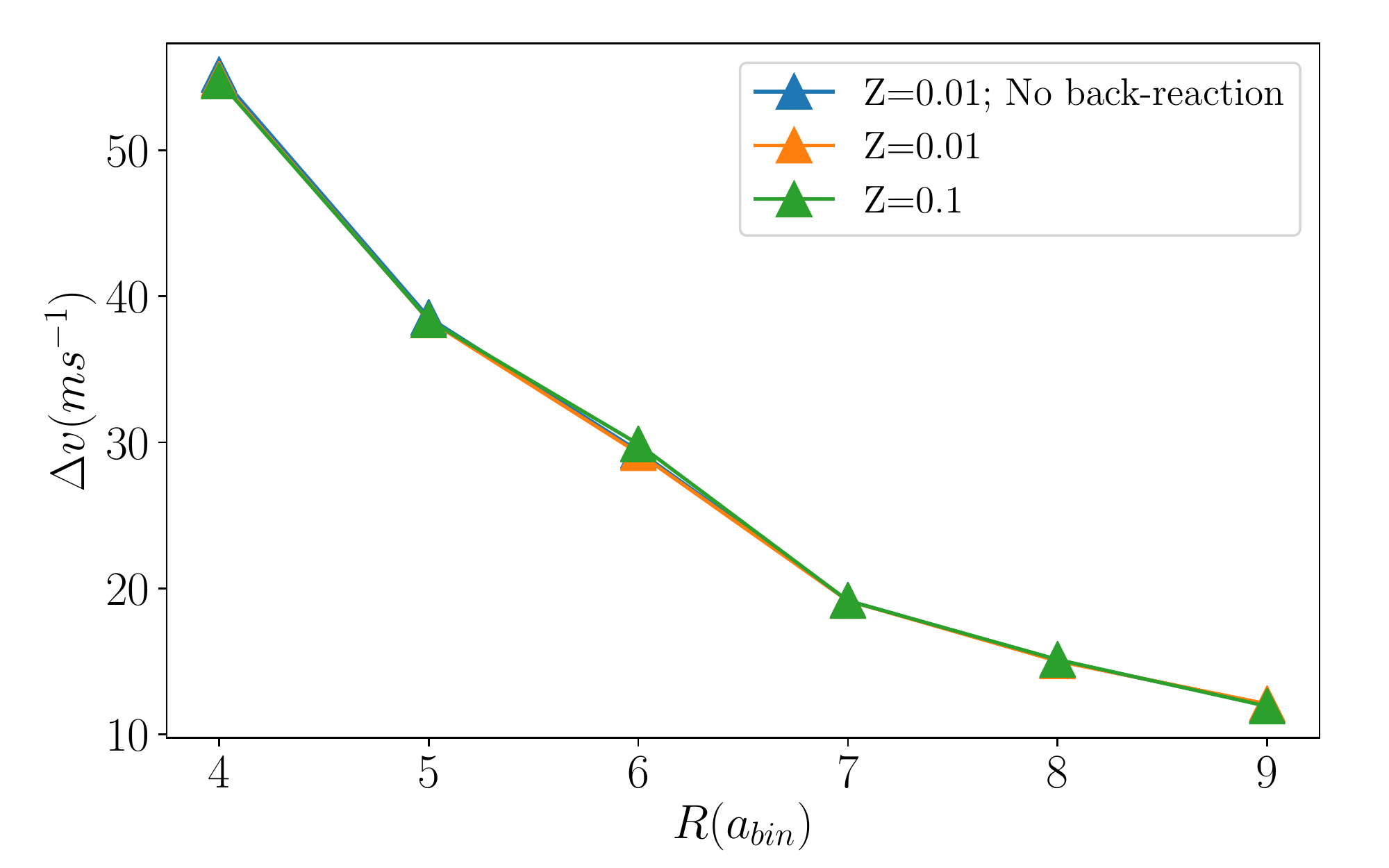}
\caption{Relative velocities between grains as a function of distance from the central binary.}
\label{fig:collisions}
\end{figure}

In order to examine the locations in the disc where turbulence may eventually lead  to collision speeds high enough to induce pebble fragmentation, we computed relative velocities between neighbouring particles as a function of distance from the central binary. For each grid cell located within a small patch of the disc $R_i\le R\le R_o$, $-0.01\le \pi/2-\theta \le 0.01$, we computed the relative velocities with respect to its neighbouring grid cells. The average of these relative velocities is then calculated for each grid cell and further time-averaged over $400$ binary orbits. The spatial distribution of collisional velocities resulting from this procedure is shown in Fig. \ref{fig:collisions}. 

For each model, the relative velocity is $\sim 40$ $m\cdot s^{-1}$ at the edge of the truncated cavity and decreases with increasing distance from the binary, reaching $\sim 11$ $ m\cdot s^{-1}$ at $R=9$ $ \abin$. These values can be compared to the  estimate of the collisional velocity $v_c$ by Ormel \& Cuzzi (2007) which predicts $v_c\approx \sqrt{3\st \alpha} c_s$ for small grains. From the right panel of Fig. \ref{fig:alpha_ecc}, we have $\alpha\sim 0.01$ at maximum and $\st\sim 0.01$ at $R=5$ $\abin$, whereas  at $R=9$ $\abin$ we have  $\alpha\sim 0.001$ and $\st \sim 0.03$. Taken together,  this leads to $v_c \sim 27$ $m\cdot s^{-1}$ and  $v_c \sim 9$ $m\cdot s^{-1}$at $R=5$ $\abin$ and $R=9$ $\abin$ respectively.

The critical velocity for dust fragmentation has been studied extensively both numerically and  experimentally (Blum \& Wurn 2008; Zsom et al. 2010; Gundlach \& Blum 2015; Wada et al. 2013; Ueda et al. 2019), and is estimated to be between $1-10$ $m\cdot s^{-1}$ for silicate particles and between $10-80$ $m\cdot s^{-1}$ for icy aggregates. However, since we are mostly interested in regions  inside the snow-line, which is approximately located at 2 AU (Hayashi 1981), we can reasonably assume that the pebbles that we consider are silicate particles that were released consequently to the sublimation of icy pebbles. Fig. \ref{fig:collisions} reveals that turbulence may lead to the disruption of those cm-sized silicate aggregates, particularly in the regions near the inner cavity. At these locations,  it would be challenging to reach dust-to-gas ratios  high enough to trigger the streaming instability. Also, the fact that collisions speeds are still higher than the fragmentation threshold velocity for a metallicity of $Z=0.1$ further suggests that the weakening of hydrodynamical turbulence due to a  local increase in the concentration of dust  would not be  significant enough to allow for the  growth of these grains. 

Close to the inner cavity, the main implication of the fragmentation of cm-sized grains would be an increase in the  population of smaller grains. However, such small grains may pass into the inner cavity and accrete onto the binary components along with the gas that flows into this region, as  the outer edge of the cavity is not expected to trap small particles efficiently.  This would tend to remove solids from  the cavity edge, resulting in a less efficient planetesimal formation through the streaming instability. The streaming instability indeed requires high solid abundances of $Z\sim 0.05$ to operate for particles with Stokes numbers of 
$\st \sim 10^{-3}$(Yang et al. 2017).  However, even if a massive planetesimal seed could be formed through the streaming instability involving these small grains, pebble accretion onto this core would be very inefficient in that case, due to the strong stirring by disc turbulence experienced by these small particles. 

Further away from the binary, we remark that the collision velocities of cm-sized grains are close to the fragmentation velocities derived from laboratory experiments. Considering distances even further than those that we include in the simulations, it is reasonable to suppose that grains could grow to sizes needed to trigger the streaming instability.

\subsection{Possible formation of dusty clumps }
Turning back to Fig. \ref{fig:2dplot}, we see from close inspection that compared to lower metallicities the case with $Z=0.1$ exhibits  dusty clumps forming in the disc midplane in the region $R\in[4,6]$. There are basically two mechanisms that may be able to lead to the emergence of such clumps. Solids might accumulate at  pressure maxima created by vortices emerging spontaneously from the hydrodynamical turbulence (Fromang \& Nelson 2005) or resulting from the Rossby Wave Instability (RWI; Lovelace et al. 1999; Li et al. 2000). Although not shown here, we examined the gas vorticity distribution and did not find any clear correlation between the vorticity and dust density distributions. An alternative possibility is that these dust clumps result from the streaming instability (Youdin \& Goodman 2005; Youdin \& Johansen 2007) that tends to enhance the concentration of solids.   Although our numerical resolution of  $\approx 15$ grid cells per scale-height  is probably insufficient to resolve the streaming instability whose most unstable wavelength is typically $\approx h H$,   recent work (Chen \& Lin 2020; Urmurhan et al. 2020) has shown that in turbulent discs with $\alpha\approx 10^{-3}$, the most unstable wavelength might be shifted toward scales as large as $H$ (Johansen et al. 2007; Umurhan et al. 2020). In that case, our numerical resolution would be sufficient to witness the SI operating. The fact that dust clumps are not formed when the particle back-reaction is neglected is a further indication of the possibly important role of the mutual dust+gas drag on the formation of these clumps.

\section{Conclusions}

In this paper, we have presented the results of  three-dimensional global hydrodynamical simulations of circumbinary discs subject to hydrodynamical turbulence driven by a parametric instability associated with the disc eccentricity (Papaloizou 2005; Barker \& Ogilvie 2014). Centimetre-sized solids modelled as a second fluid are introduced and we study the impact of turbulence on dust vertical settling. We also compare models with and without the effect of the particle back-reaction onto the gas included, and examine the impact of varying  the solid abundance. 

In agreement with our previous findings (Pierens et al. 2020), the particle vertical profile reaches a quasi-stationary state once turbulent diffusion counterbalances gravitational settling. At  steady-state, we find that a higher solid-to-gas ratio leads to a smaller dust scale height. This is a direct consequence of the effect of  dust back-reaction onto the gas, which induces smaller vertical velocity fluctuations in the gas, resulting in a smaller dust vertical diffusion coefficient. 

In the vicinity of the inner cavity formed by the binary, cm-sized grains are significantly stirred by the turbulence, such that the dust and gas components have similar scale-heights. 
This renders pebble accretion inefficient in growing  low-mass planetary cores, as the accretion rate is reduced by a factor of $\racc/H_d$ compared to a laminar disc. Although increasing the dust-to-gas ratio slightly favours dust settling, we find that growing a Ceres-mass object to a 10 $M_\oplus$ core in  a disc with metallicity $Z=0.1$ still requires a time significantly longer than the disc lifetime. Growing such a planetary core, however, be possible at large distances from the binary, where the turbulence is weak enough so that pebble accretion onto a 1 $M_\oplus$ planet becomes an essentially 2D process. 

We also examined, in the different regions of the disc,  how the impact velocities induced by the turbulence compare with the typical fragmentation velocities of silicate aggregates. Collision velocities are found to be significantly higher than the critical disruption velocity close to the inner binary, whereas further away from the binary these are comparable to the fragmentation velocity. This suggests the possibility of grain growth at large distances from  the binary to sizes sufficient to trigger the streaming instability at moderate solid-to-gas ratios. Near the inner cavity, fragmentation would rather lead to an increase in the abundance of grains that have mm-sizes and smaller. However, it seems unlikely that such small grains can participate in the streaming instability because: i) these are expected to pass through the edge of the inner cavity, and ii) a high dust-to-gas ratio is required to trigger a streaming instability involving mm-sized grains.

Taken together, these results suggest that forming a circumbinary planet, such as Kepler-16b, in-situ at the cavity edge, through a scenario that combines the streaming instability to build a seed planetesimal, followed by pebble accretion onto this seed object to form a planetary core that accretes gas, is very difficult to achieve. We simply note here that Kepler-16b orbits at a distance equivalent to $3.142 a_{\rm bin}$ (Doyle et al. 2011), placing it right at the inner edge of the disc where the disturbance due to the binary and the hydrodynamical turbulence is at its strongest. This suggests that a more realistic scenario would involve formation of the planet further from the binary where the streaming instability and pebble accretion can operate more efficiently due to the lower levels of turbulence, followed by migration to the edge of the tidally truncated cavity formed by the binary. 
\section*{Acknowledgments}
Computer time for this study was provided by the computing facilities MCIA (M\'esocentre de Calcul Intensif Aquitain) of the Universite de Bordeaux and by HPC resources of Cines under the allocation  A0090406957 made by GENCI (Grand Equipement National de Calcul Intensif).  CPM and RPN acknowledge support from STFC through grants ST/P000592/1 and ST/T000341/1.

\section*{Data Availability}

The data underlying this article will be shared on reasonable request to the corresponding author.


\begin{thebibliography}{}
\bibitem[Artymowicz, \& Lubow(1994)]{1994ApJ...421..651A} Artymowicz, P., \& Lubow, S.~H.\ 1994, ApJ, 421, 651
\bibitem[Bae et al.(2016)]{2016ApJ...829...13B} Bae, J., Nelson, R.~P., Hartmann, L., et al.\ 2016, ApJ, 829, 13
\bibitem[Barker, \& Ogilvie(2014)]{2014MNRAS.445.2637B} Barker, A.~J., \& Ogilvie, G.~I.\ 2014, MNRAS, 445, 2637
\bibitem[Ben{\'\i}tez-Llambay, \& Masset(2016)]{2016ApJS..223...11B} Ben{\'\i}tez-Llambay, P., \& Masset, F.~S.\ 2016, ApJS, 223, 11
\bibitem[Blum \& Wurm(2008)]{2008ARA&A..46...21B} Blum, J. \& Wurm, G.\ 2008, ARAA, 46, 21. doi:10.1146/annurev.astro.46.060407.145152
\bibitem[Chen \& Lin(2020)]{2020ApJ...891..132C} Chen, K. \& Lin, M.-K.\ 2020, ApJ, 891, 132. doi:10.3847/1538-4357/ab76ca
\bibitem[Doyle et al.(2011)]{2011Sci...333.1602D} Doyle, L.~R., Carter, J.~A., Fabrycky, D.~C., et al.\ 2011, Science, 333, 1602. doi:10.1126/science.1210923
\bibitem[Dr{\k{a}}{\.z}kowska \& Dullemond(2014)]{2014A&A...572A..78D} Dr{\k{a}}{\.z}kowska, J. \& Dullemond, C.~P.\ 2014, A\& A, 572, A78. doi:10.1051/0004-6361/201424809
\bibitem[Dubrulle et al.(1995)]{1995Icar..114..237D} Dubrulle, B., Morfill, G., \& Sterzik, M.\ 1995, Icarus, 114, 237
\bibitem[Hayashi(1981)]{1981PThPS..70...35H} Hayashi, C.\ 1981, Progress of Theoretical Physics Supplement, 70, 35
\bibitem[Fromang \& Nelson(2005)]{2005MNRAS.364L..81F} Fromang, S. \& Nelson, R.~P.\ 2005, MNRAS, 364, L81. doi:10.1111/j.1745-3933.2005.00109.x
\bibitem[Fromang \& Papaloizou(2006)]{2006A&A...452..751F} Fromang, S. \& Papaloizou, J.\ 2006, A\& A, 452, 751. doi:10.1051/0004-6361:20054612
\bibitem[Gundlach \& Blum(2015)]{2015ApJ...798...34G} Gundlach, B. \& Blum, J.\ 2015, ApJ, 798, 34. doi:10.1088/0004-637X/798/1/34
\bibitem[Johansen et al.(2007)]{2007Natur.448.1022J} Johansen, A., Oishi, J.~S., Mac Low, M.-M., et al.\ 2007, Nature, 448, 1022. doi:10.1038/nature06086
\bibitem[Klahr \& Schreiber(2020)]{2020ApJ...901...54K} Klahr, H. \& Schreiber, A.\ 2020, ApJ, 901, 54. doi:10.3847/1538-4357/abac58
\bibitem[Kley \& Haghighipour(2014)]{2014A&A...564A..72K} Kley, W., \& Haghighipour, N.\ 2014, A\& A, 564, A72
\bibitem[Kley, \& Haghighipour(2015)]{2015A&A...581A..20K} Kley, W., \& Haghighipour, N.\ 2015, A\& A, 581, A20
\bibitem[Kostov et al.(2016)]{2016ApJ...827...86K} Kostov, V.~B., Orosz, J.~A., Welsh, W.~F., et al.\ 2016, ApJ, 827, 86
\bibitem[Li et al.(2000)]{2000ApJ...533.1023L} Li, H., Finn, J.~M., Lovelace, R.~V.~E., \& Colgate, S.~A.\ 2000, ApJ, 533, 1023 
\bibitem[Lin(2019)]{2019MNRAS.485.5221L} Lin, M.-K.\ 2019, MNRAS, 485, 5221. doi:10.1093/mnras/stz701
\bibitem[Lin \& Youdin(2017)]{2017ApJ...849..129L} Lin, M.-K. \& Youdin, A.~N.\ 2017, ApJ, 849, 129. doi:10.3847/1538-4357/aa92cd
\bibitem[Lovelace et al.(1999)]{1999ApJ...513..805L} Lovelace, R.~V.~E., Li, H., Colgate, S.~A., \& Nelson, A.~F.\ 1999, ApJ, 513, 805 
\bibitem[Ormel \& Cuzzi(2007)]{2007A&A...466..413O} Ormel, C.~W. \& Cuzzi, J.~N.\ 2007, A\& A, 466, 413. doi:10.1051/0004-6361:20066899
\bibitem[Meschiari(2012)]{2012ApJ...752...71M} Meschiari, S.\ 2012, ApJ, 752, 71
\bibitem[Meschiari(2012)]{2012ApJ...761L...7M} Meschiari, S.\ 2012, ApJL, 761, L7
\bibitem[Mutter et al.(2017)]{2017MNRAS.465.4735M} Mutter, M.~M., Pierens, A., \& Nelson, R.~P.\ 2017, MNRAS, 465, 4735
\bibitem[Nelson et al.(2013)]{2013MNRAS.435.2610N} Nelson, R.~P., Gressel, O., \& Umurhan, O.~M.\ 2013, MNRAS, 435, 2610
\bibitem[Paardekooper et al.(2012)]{2012ApJ...754L..16P} Paardekooper, S.-J., Leinhardt, Z.~M., Th{\'e}bault, P., \& Baruteau, C.\ 2012, ApJL, 754, L16
\bibitem[Papaloizou(2005)]{2005A&A...432..743P} Papaloizou, J.~C.~B.\ 2005a, A \& A, 432, 743
\bibitem[Papaloizou(2005)]{2005A&A...432..757P} Papaloizou, J.~C.~B.\ 2005b, A\& A, 432, 757
\bibitem[Penzlin(2021)]{2021A&A...645A..68P} Penzlin, A., Kley, W., Nelson, R.~P. \ 2021, A\&A, 645, 68 
\bibitem[Pierens, \& Nelson(2007)]{2007A&A...472..993P} Pierens, A., \& Nelson, R.~P.\ 2007, A\& A, 472, 993
\bibitem[Pierens, \& Nelson(2008)]{2008A&A...478..939P} Pierens, A., \& Nelson, R.~P.\ 2008, A\& A, 478, 939
\bibitem[Pierens \& Nelson(2008)]{2008A&A...483..633P} Pierens, A., \& Nelson, R.~P.\ 2008, A\&A, 483, 633
\bibitem[Pierens \& Nelson(2013)]{2013A&A...556A.134P} Pierens, A., \& Nelson, R.~P.\ 2013, A\& A, 556, A134
\bibitem[Pierens et al.(2020)]{2020MNRAS.496.2849P} Pierens, A., McNally, C.~P., \& Nelson, R.~P.\ 2020, MNRAS, 496, 2849. doi:10.1093/mnras/staa1550
\bibitem[Riols \& Lesur(2018)]{2018A&A...617A.117R} Riols, A. \& Lesur, G.\ 2018, A\& A, 617, A117. doi:10.1051/0004-6361/201833212
\bibitem[Sch{\"a}fer et al.(2017)]{2017A&A...597A..69S} Sch{\"a}fer, U., Yang, C.-C., \& Johansen, A.\ 2017, A\& A, 597, A69. doi:10.1051/0004-6361/201629561
\bibitem[Stoll et al.(2017)]{2017A&A...599L...6S} Stoll, M.~H.~R., Kley, W., \& Picogna, G.\ 2017, A\&A, 599, L6. doi:10.1051/0004-6361/201630226
\bibitem[Ueda et al.(2019)]{2019ApJ...871...10U} Ueda, T., Flock, M., \& Okuzumi, S.\ 2019, ApJ, 871, 10. doi:10.3847/1538-4357/aaf3a1
\bibitem[Umurhan et al.(2020)]{2020ApJ...895....4U} Umurhan, O.~M., Estrada, P.~R., \& Cuzzi, J.~N.\ 2020, ApJ, 895, 4. doi:10.3847/1538-4357/ab899d
\bibitem[Wada et al.(2013)]{2013A&A...559A..62W} Wada, K., Tanaka, H., Okuzumi, S., et al.\ 2013, A\& A, 559, A62. doi:10.1051/0004-6361/201322259
\bibitem[Wienkers, \& Ogilvie(2018)]{2018MNRAS.477.4838W} Wienkers, A.~F., \& Ogilvie, G.~I.\ 2018, MNRAS, 477, 4838
\bibitem[Yang et al.(2009)]{2009ApJ...707.1233Y} Yang, C.-C., Mac Low, M.-M., \& Menou, K.\ 2009, ApJ, 707, 1233
\bibitem[Yang et al.(2017)]{2017A&A...606A..80Y} Yang, C.-C., Johansen, A., \& Carrera, D.\ 2017, A\& A, 606, A80. doi:10.1051/0004-6361/201630106
\bibitem[Youdin \& Goodman(2005)]{2005ApJ...620..459Y} Youdin, A.~N. \& Goodman, J.\ 2005, ApJ, 620, 459. doi:10.1086/426895
\bibitem[Youdin \& Johansen(2007)]{2007ApJ...662..613Y} Youdin, A. \& Johansen, A.\ 2007, ApJ, 662, 613. doi:10.1086/516729
\bibitem[Youdin, \& Lithwick(2007)]{2007Icar..192..588Y} Youdin, A.~N., \& Lithwick, Y.\ 2007, Icarus, 192, 588
\bibitem[Zhu et al.(2009)]{2009ApJ...694.1045Z} Zhu, Z., Hartmann, L., \& Gammie, C.\ 2009, ApJ, 694, 1045. doi:10.1088/0004-637X/694/2/1045
\bibitem[Zsom et al.(2010)]{2010A&A...513A..57Z} Zsom, A., Ormel, C.~W., G{\"u}ttler, C., et al.\ 2010, A\& A, 513, A57. doi:10.1051/0004-6361/200912976
\end{thebibliography}
\end{document}